\documentclass[fleqn,usenatbib]{mnras}

\usepackage[T1]{fontenc}

\DeclareRobustCommand{\VAN}[3]{#2}
\let\VANthebibliography\thebibliography
\def\thebibliography{\DeclareRobustCommand{\VAN}[3]{##3}\VANthebibliography}

\usepackage{graphicx}	% Including figure files
\usepackage{amsmath}	% Advanced maths commands
\usepackage{amssymb}	% Extra maths symbols

\usepackage{newtxtext,newtxmath}

\title[Binaries and IMBHs]{Using Binaries in Globular Clusters to Catch Sight of Intermediate-Mass Black Holes}

\author[Aros et al.]{
Francisco I. Aros$^{1,2}$\thanks{E-mail: francisco.aros@univie.ac.at}, Anna C. Sippel$^{2}$, Alessandra Mastrobuono-Battisti$^{3,4}$, 
\newauthor Paolo Bianchini$^{5}$, Abbas Askar$^{3}$, Glenn van de Ven$^{1}$ 
\\
$^{1}$Department of Astrophysics, University of Vienna, T\"urkenschanzstrasse 17, 1180 Vienna, Austria\\
$^{2}$Max Planck Institute for Astronomy, K\"onigstuhl 17, D-69117 Heidelberg, Germany\\
$^{3}$Lund Observatory, Department of Astronomy and Theoretical Physics, Lund University, Box 43, SE-221 00 Lund, Sweden\\
$^{4}$GEPI, Observatoire de Paris, PSL Research University, CNRS, Place Jules Janssen, 92190 Meudon, France\\
$^{5}$Observatoire Astronomique de Strasbourg, Universit\'e de Strasbourg, CNRS UMR7550, Strasbourg, France\\
}

\date{Accepted XXX. Received YYY; in original form ZZZ}

\pubyear{2021}

\begin{document}
\label{firstpage}
\pagerange{\pageref{firstpage}--\pageref{lastpage}}
\maketitle

\begin{abstract}
The dynamical evolution of globular clusters (GCs) is tied to their binary population, as binaries segregate to the cluster centre, leading to an increased binary fraction in the core. This central overabundance of mainly hard binaries can serve as a source of energy for the cluster and has a significant effect on the observed kinematics, such as artificially increasing the observed line-of-sight velocity dispersion. 

We analyse the binary fractions and distributions of 95 simulated GCs, with and without an intermediate-mass black hole (IMBH) in their centre. We show that an IMBH will not only halt the segregation of binaries towards the cluster centre, but also, directly and indirectly, disrupt the binaries that segregate, thus depleting binaries in the cluster core. We illustrate this by showing that clusters with an IMBH have fewer binaries and flatter radial binary distributions than their counterparts without one. These differences in the binary fraction and distribution provide an additional indicator for the presence of a central IMBH in GCs. In addition, we analyse the effects of the binary fraction on the line-of-sight velocity dispersion in the simulated GCs and find that binaries can cause an overestimation of up to $70\%$ of the velocity dispersion within the core radius. Using recent VLT/MUSE observations of NGC 3201 by \protect\cite{giesers_2019}, we find an overestimation of $32.2\pm7.8\%$ in the velocity dispersion that is consistent with the simulations and illustrates the importance of accurately accounting for the binary population when performing kinematic or dynamical analysis.
\end{abstract}


%%%%%%%%%%%%%%%%%%%%%%%%%%%%%%%%%%%%%%%%%%%%%%%%%%

%%%%%%%%%%%%%%%%% BODY OF PAPER %%%%%%%%%%%%%%%%%%

\section{Introduction}

Intermediate-mass black holes (IMBHs) with masses of $10^2-10^5\, M_{\odot}$ are one of the missing links in the formation and growth of the supermassive black holes found at the centres of massive galaxies, as early formed IMBHs can serve as seeds capable of growing up to the masses of supermassive black holes supporting the quasars observed at high-redshift \citep[e.g., ][]{haiman_2013}. The possible formation scenarios of IMBHs \citep[e.g., ][]{portegies_zwart_2004,giersz_2015,gonzalez_2021,rizzuto_2021} point to dense stellar systems as their place of origin and therefore many studies have searched for IMBHs in the centres of globular clusters.

Globular clusters (GCs) are dense stellar systems with up to a few million stars and half-light radii of 2-5 pc \citep[][2010 edition]{harris_1996}. In the last two decades, many studies have looked at the centres of Galactic GCs to search for kinematic evidence for the presence of an IMBH \citep[][to name a few]{noyola_2008,van_der_marel_2010,mcnamara_2012,lutzgendorf_2013,lanzoni_2013,kamann_2014}. However, limitations on the observed kinematics \citep{de_vita_2017} and dynamical modelling \citep{aros_2020} may hinder the robust detection of an IMBH. No clear evidence for an IMBH has been found so far within Galactic GCs. The possible presence of a black hole system (BHS, see Section \ref{sec:imbh_bin}) in the cluster centre adds to this complex scenario, as it can produce similar velocity dispersions in the cluster core as a central IMBH \citep{zocchi_2019,baumgardt_2019,mann_2019,vitral_2021}.

Binary stars play a crucial role in the evolution of GCs. The dynamical evolution of GCs is driven by two-body relaxation that triggers mass segregation \citep{spitzer_1987}. Binaries, being more massive than single stars on average (two stars instead of one), segregate earlier to the cluster centre. Simulations of GCs with primordial binaries \citep{heggie_2006,hurley_2007,chatterjee_2010,wang_2016} show how binaries segregate towards the cluster centre while becoming harder or getting disrupted by encounters with other stars in the dense core. These two processes provide energy to the cluster and play a significant role during the core-collapse period of the GCs’ evolution.

Observations of binaries in Galactic GCs show that binaries follow the behaviour described in simulations. Measurements of the binary fraction at different radii show a decreasing gradient with radius \citep[see][]{sollima_2007,milone_2012,ji_2015} and serve as proof for the mass segregation of binaries in GCs. The connection between binaries and the dynamical evolution of the GC has also been explored with specific types of binaries and products of binary evolution. \cite{ferraro_2012, ferraro_2018} used the distribution of blue-straggler stars (BSS) to study the dynamical age of GCs. As formation channels of BSS include binary interactions \citep{mapelli_2006,chatterjee_2013}, the tracing of BSS stars is linked to the overall process of mass segregation. In a similar way, the distribution of X-ray binaries in GCs also describes the same process \citep{Cheng_2019a,cheng_2019b}. The mass segregation driven by two body relaxation not only brings binary systems closer to the cluster centre, but it also efficiently accumulates more massive stars and massive stellar remnants such as stellar mass black holes.

A massive object in the cluster centre such as an IMBH will alter the dynamical evolution of the GC. The IMBH hampers mass segregation by working as a source of energy for the cluster \citep{baumgardt_2004,trenti_2007,gill_2018} and in the process altering the distribution of binaries within the cluster core. Moreover, simulations of GCs with a central IMBH and primordial binaries show that binaries in the core are disrupted more efficiently than in the case without an IMBH \citep{trenti_2007}. Whereas this is more likely to happen due to the high density of stars near the IMBH, some binaries could strongly interact with the IMBH. The strong interaction with the IMBH breaks the binary producing high-velocity stars \cite[e.g., ][]{hills_1988,fragione_2019,subr_2019}.
 
In this work, we study the effects that the presence of an IMBH has on the binary population of its host GC. Motivated by the recent observations of binaries in NGC 3201 with VLT/MUSE by \cite{giesers_2019}, we use a sample of simulated GCs from the \textsc{MOCCA}-Survey database I \citep{askar_2017a} with two goals: (1) explore the contamination of binaries in the line-of-sight velocity dispersion and (2) use the detected binary sample to study the binary fraction in GCs with and without an IMBH. The first point is effectively extending the discussion of Figure 4 in \cite{aros_2020} where we show that binaries can systematically increase the observed line-of-sight velocity dispersion.

In Section \ref{sec:methods} we describe the sample of simulated GCs and the detection method for binary stars. In Section \ref{sec:kin_bin} we explore the contamination of binaries in the observed line-of-sight velocity dispersion and its implication for current observations. In Section \ref{sec:imbh_bin} we analyze the relation of binaries with the central IMBH and discuss how we could use their co-evolution as an indication for the presence of an IMBH. Finally, in Section \ref{sec:summary} we summarize our findings and motivate future work.

\section{Simulations and binary identification}
\label{sec:methods}
The dynamical evolution of GCs is determined by two-body interactions leading towards partial energy equipartition \citep[see e.g. ][]{spitzer_1969,trenti_2013,bianchini_2016b}. A consequence of the drive towards energy equipartition is mass segregation, where more massive stars sink towards the centre of the GC. Binaries being on average more massive than single stars segregate faster to the centre, ultimately increasing the binary fraction towards the cluster core. To study the general behaviour of binaries under the presence of an IMBH, we identify binaries in mock data from simulated GCs as described in the following sections.

\subsection{Simulations and mock data}
\label{sec:simulations}
We have selected a sample of 284 simulated clusters from the \textsc{MOCCA}-Survey Database I \citep{askar_2017a} to study the effects of binaries in the observed kinematics and the interaction of binaries with a central IMBH. These simulations were evolved to $12\,\text{Gyr}$ using the \textsc{MOCCA} code \citep{hypki_2013,giersz_2013}, which follows a state-of-the-art implementation of the Monte Carlo method first proposed by \cite{henon_1971b,henon_1971a}. All 284 simulations have different initial conditions (see Appendix \ref{secA:initial_condition}), but share the same initial binary fraction $f_{\text{bin}}=10\%$, consistent with observations \citep{sollima_2007,milone_2012}.   

As described in detail by \cite{giersz_2015}, IMBHs in \textsc{MOCCA} simulations form via dynamical interactions in two kinds of scenarios. The “FAST” scenario starts from the beginning of the simulation and requires an extremely high initial central density \cite[$\gtrsim10^6\,M_{\odot}/\text{pc}^3$, see also][]{hong_2020}, where a BH system can form early in the star cluster evolution and drive the formation of an IMBH by the dynamical interactions of single and binary BHs \citep[similar to the runaway collapse proposed by][]{portegies_zwart_2004}.  The “SLOW” scenario, on the other hand, happens at later stages in the cluster evolution and in less dense systems (with central densities of $10^5\,M_{\odot}/\text{pc}^3$). The mass of a single BH can grow through dynamical interactions and mass accretion from a binary companion.

The binary-star evolution code  \citep[\textsc{BSE},][]{hurley_2002} models the binary stellar evolution in MOCCA, while the \textsc{FEWBODY} code \citep{fregeau_2004} drives the dynamical interactions of binaries with single or binary stars. The interplay of both codes allows \textsc{MOCCA} to follow the evolution of binary stars in a realistic way, while also keeping short simulation times. \cite{hypki_2013} describes the interplay of both codes within \textsc{MOCCA} extensively.

\textsc{MOCCA} simulations are spherically symmetric and only have three coordinates for each single star and binary system: the radial position $r$, and the radial $\text{v}_r$ and tangential $\text{v}_t$ velocities. Therefore, we project each simulated GC into cartesian coordinates by randomly sampling the missing coordinates.  We then project the simulated GC into the sky to receive the position of each single and binary star, along with their radial velocity (RV) along the line-of-sight and the proper motions. We follow the same projection as in \cite{aros_2020} (see their Figure 1). For binaries, we consider the centre of mass and we follow the same procedure.

For every binary system in the cluster, we obtain each component’s relative velocity with respect to their centre of mass, drawing the orbit from the known eccentricity, semi-major axis and masses of each star in the binary. From these orbital parameters, we can obtain the individual positions and velocities of both components of the binary, the period and angular momentum. We randomly select the current position of the binary by sampling a random time between zero and one orbital period. In the same way, we randomly orientate the binary’s orbital plane. Whereas this approach is similar to the projection described by \cite{askar_2018}, we use the direct solution of the Kepler orbit rather than the parameterization of the orbit using the eccentric anomaly.

We generate mock data for each cluster by adding errors to the velocities that are consistent with current observations. For line-of-sight radial velocities, we use the error distribution by magnitude from MUSE data \citep{giesers_2019}, with a median value of $3\,\text{km/s}$ at $m_{\text{V}}\sim 18\,\text{mag}$ at a heliocentric distance of $\sim5\,\text{kpc}$ (see Figure \ref{fig:vz_error}). For proper motions, we use the projected velocities as given on the simulation (for binary systems we use the centre of mass velocity), and assume the median error from \cite{libralato_2018}, which is $0.1\,\text{mas/yr}$, and that corresponds to $2\,\text{km/s}$ at $5\,\text{kpc}$ (see also Appendix \ref{secA:errors}). In both cases, we add noise to the simulation’s velocities by sampling the noise from a Gaussian distribution centred in zero with a dispersion given by the assigned error. Whereas we have assumed that all the clusters are at a heliocentric distance of $5\,\text{kpc}$ to have a comparable sample of stellar masses and observed stars, the simulated clusters have different Galactocentric distances ranging within $1-15\,\text{kpc}$. 

We have limited the “observed” stars within each cluster to those with line-of-sight RV errors smaller than $3\,\text{km/s}$. This set a lower limit in magnitude around $1\,\text{mag}$ below the main-sequence-turn-off (or $\sim 18\,\text{mag}$), similar as used by \cite{aros_2020} and motivated by current observations \citep{libralato_2018, giesers_2019}. From the sample of 284 simulated GCs in the \textsc{MOCCA}-Survey database I with an initial binary fraction of $10\%$, we selected a subsample of clusters that have: (a) more than $1000$ stellar systems (binary and single) within the selected sample of sufficiently bright stars, and (b) an intrinsic velocity dispersion higher than $4\,\text{km/s}$, which is twice the median velocity error in line-of-sight velocities and proper motions at the assumed distance of $5\,\text{kpc}$. These two criteria help to reduce the stochasticity due to low numbers (binary fraction), and to measure velocity dispersion values that are not dominated by observational errors. Our sample then decreases to 95 GCs.

\begin{figure}
    \centering
    \includegraphics[width=0.9\linewidth]{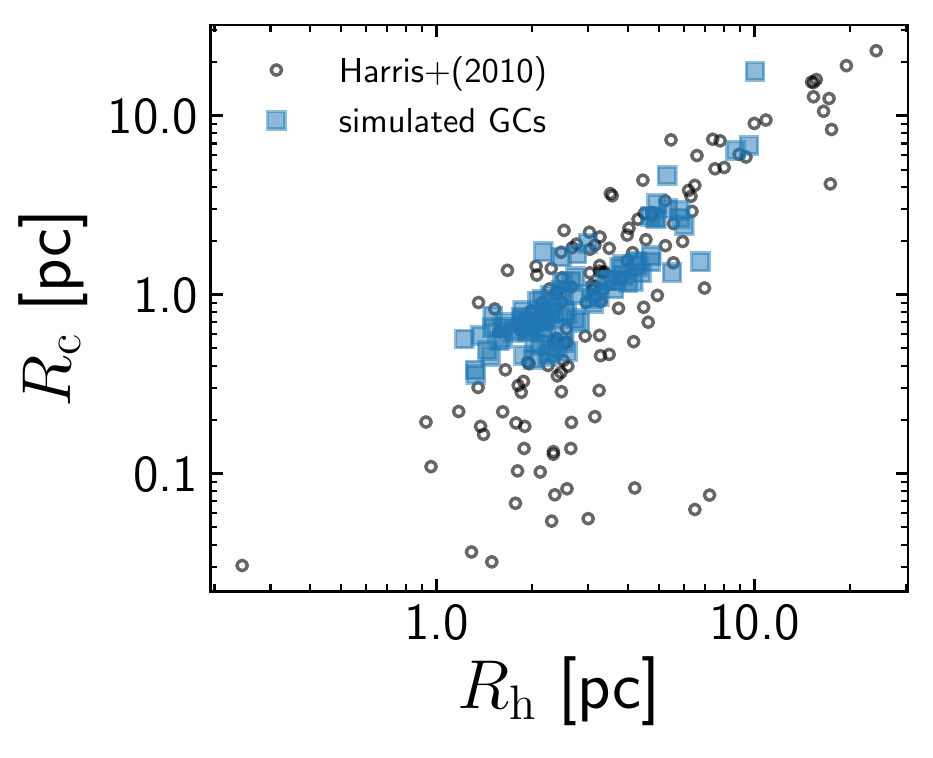}
    \caption{Core $R_{\text{c}}$ and half-light $R_{\text{h}}$ radii for simulated GCs. We have selected a sample of 95 simulated GCs with and without a central IMBH, with initial binary fraction of $10\%$ (light-blue squares) and additional requirements on the quality of the data as described in Section \protect\ref{sec:simulations}. As a reference we show Galactic GCs from \protect\cite{harris_1996} (2010 edition, as empty circles). Our selected sample has comparable sizes as real GCs.}
    \label{fig:sample_comparison}
\end{figure}

In the following sections of this work, we refer to the projected core ($R_{\text{c}}$) and half-light radii ($R_{\text{h}}$) of each cluster. Both radii are extracted from \cite{arca_sedda_2019}, who fitted a King profile \citep{king_1962} to the cumulative surface brightness profile, following the method described by \cite{morscher_2015}. Figure \ref{fig:sample_comparison} shows the core and half-light radii for the subsample of 95 simulated GCs and for Galactic GCs from \cite{harris_1996} (2010 edition). The sizes of our selected sample are comparable to observed GCs.

\subsection{Identification of binaries}
\label{sec:bin_det}

We identify binaries in the mock data by measuring the observed RV for each star at different observational epochs, i.e. different observations with time scales of hours, weeks or months\footnote{ We use 19 epochs given by $t_{\text{epoch}}=[$0, 6, 15, 24, 33, 360, 375, 384, 384.03, 384.06, 384.09, 384.12, 384.15, 384.93, 384.96, 384.99, 385.02, 385.05, 393$]$ days.}. We do not evolve the simulations between epochs; therefore, the positions and velocities of single stars, and the centre of mass position and velocity of binary systems remain fixed\footnote{Single stars and the centre of mass of binary systems would not significantly move in this time range since it is much smaller than the dynamical time.}. Any variation on the RV of single stars is only due to observational errors. For binary stars, while the center of mass remains fixed, we update the position and velocities of both components at each epoch. We use the orbital parameters and subsequent orbit to follow the binary components relative positions and velocities around the binary centre of mass. The “observed” RV for the binary star is given by the luminosity weighted velocity along the line of sight:
\begin{equation}
    \text{v}_{\text{los}} = \text{v}_{\text{cm}} + \frac{\text{v}_{1} L_{1} + \text{v}_{2}L_{2}}{L_1+L_2}\,,\label{eq:obs_los}
\end{equation}
where $\text{v}_{\text{cm}}$ is the centre of mass velocity of the binary, and $\text{v}_i$ and $L_i$ are the velocity and luminosity of each component. 

Once we calculate the RVs for all epochs for single and binary stars in the simulated cluster, we proceed to assign a probability of variability. To do so, we follow the approach of \cite{giesers_2019}, described in the following. The probability of variability is defined by analysing the RV curve and the scatter around the mean velocity for all epochs. For each star with $n$ different epochs, the scatter in RV can be described by 
\begin{equation}
\chi^2_{\text{obs}} = \sum_{i=0}^n\frac{(\text{v}_i-\overline{\text{v}})^2}{\delta \text{v}_i^2}\,.\label{eq:chi2}
\end{equation}

In a GC with only single stars, the scatter is dominated by errors only, and the distribution of observed $\chi^2$ resembles a theoretical $\chi^2$ distribution with $\nu=n-1$ degrees of freedom. Variability due to the motion of binaries would show as an extended tail in the observed $\chi^2$ distribution or as a change in the slope of the cumulative distribution, because the scatter will be dominated by the orbital motions rather than observational errors. 

Using this idea, \cite{giesers_2019} define the probability of a star to have a variable RV as
\begin{equation}
    P(\chi^2_i,\nu_i) = \frac{F(\chi^2_i,\nu_i)_{\text{theo}}-F(\chi^2_i,\nu_i)_{\text{obs}}}{1 - F(\chi^2_i,\nu_i)_{\text{obs}}}\,,\label{eq:prob}
\end{equation}
where $F(\chi_i^2,\nu_i)$ is the cumulative $\chi^2$ distribution given $\nu_i$ degrees of freedom, evaluated at the measured $\chi^2$ value of the $i$-th star. In our case, we use the same number of epochs for each star and cluster, a version which considers a different number of epochs is described in \cite{giesers_2019}. In the case of the simulated clusters, we can link the probability of variable RV to the probability of being a binary star, as no other effect could produce the same signature; therefore, we will use $P_{\text{bin}} = P(\chi^2_i,\nu_i)$. 

\begin{figure*}
    \centering
    \includegraphics[width=0.8\linewidth]{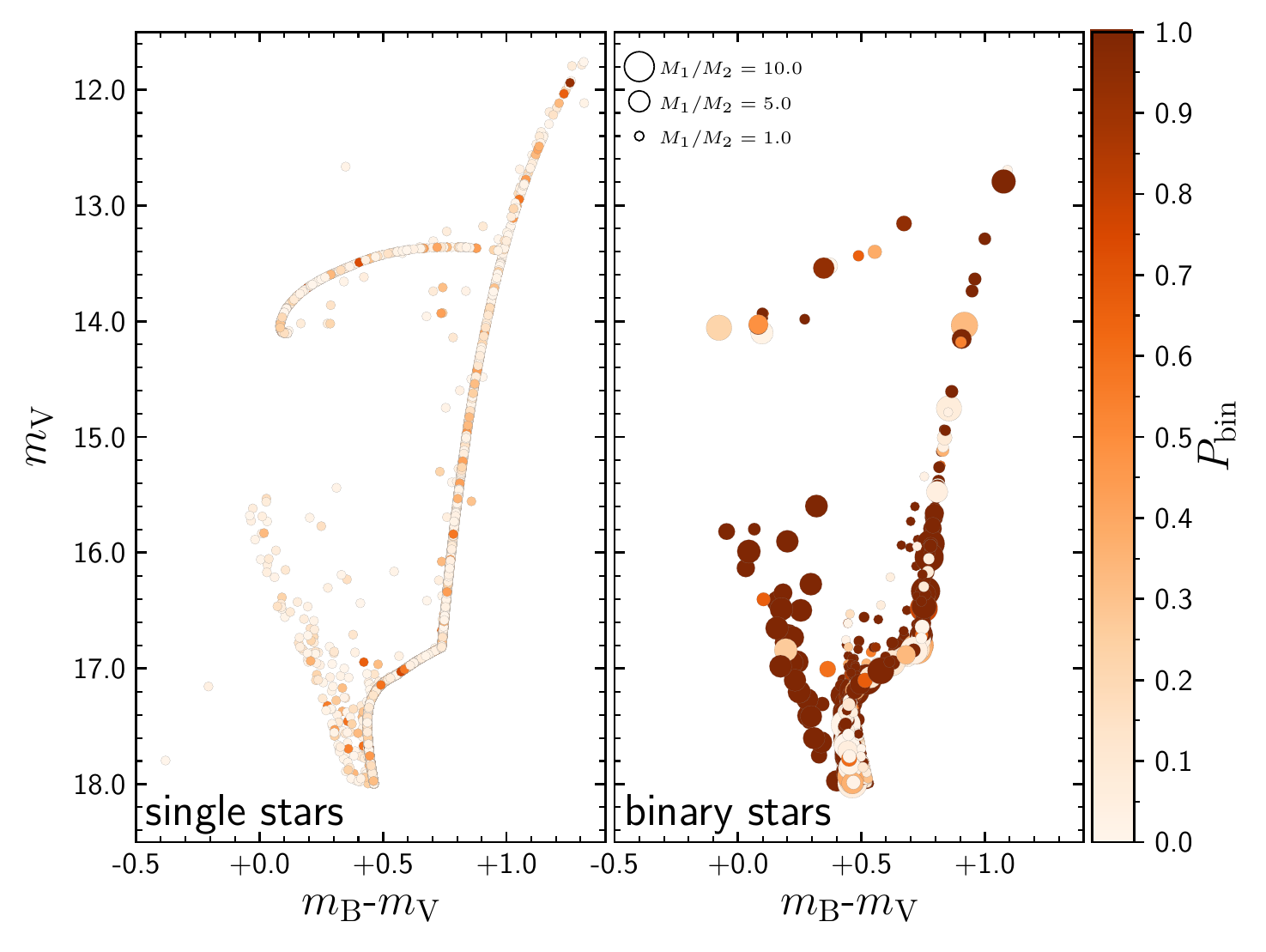}
    \caption{Color-magnitude diagram for single (left panel) and binary (right panel) stars in a simulated GC at $12\,\text{Gyr}$. The points are colour coded by the probability of being a binary with lighter colours corresponding to single stars. We can see in the left panel that most single stars have a low probability for being a binary. On the other hand, in the right panel not all binaries are identified, this is due to either having long periods or very face-on orbital planes. The size of the symbols on the right panel indicates the mass ratio between the binary components. We do not observe any particular trend with magnitude for the binary detectability. Binaries in the blue-straggler branch are likely to be detected (close binaries).}
    \label{fig:cmd_bin}
\end{figure*}

Figure \ref{fig:cmd_bin} shows the colour-magnitude diagram for a simulated GC with a binary fraction of $f_{\text{bin}}=7.8\%$ at $12\,\text{Gyr}$\. This cluster does not have a central IMBH, and neither a significant amount of retained stellar-mass black holes, and represents an average cluster. In a previous work, we built dynamical models for it to study the limitations of dynamical models to measure the mass profile of GCs and detect a central IMBH \citep[see the model named "no IMBH/BHS" in ][]{aros_2020}. Each star is colour coded by its probability of being variable in RV given by $P_{\text{bin}}$. The left panel shows all single stars in our luminosity selection, the large majority of single stars have a low probability of being variable in RV and are unlikely to be binaries. On the right panel of Figure \ref{fig:cmd_bin}, we show all binaries in the simulated cluster also colour coded by their probability of being variable in RV, the sizes of the symbols represent the mass ratio between components. While many binaries have indeed a high probability of being variable in RV, a significant fraction does not and therefore are not identified as binaries. If the velocity amplitude of the binary in the RV curve is not large enough, it will not show as having a high scatter in $\chi^2$ and will be assigned a low probability of variability in RV. 

\begin{figure}
    \centering
    \includegraphics[width=0.9\linewidth]{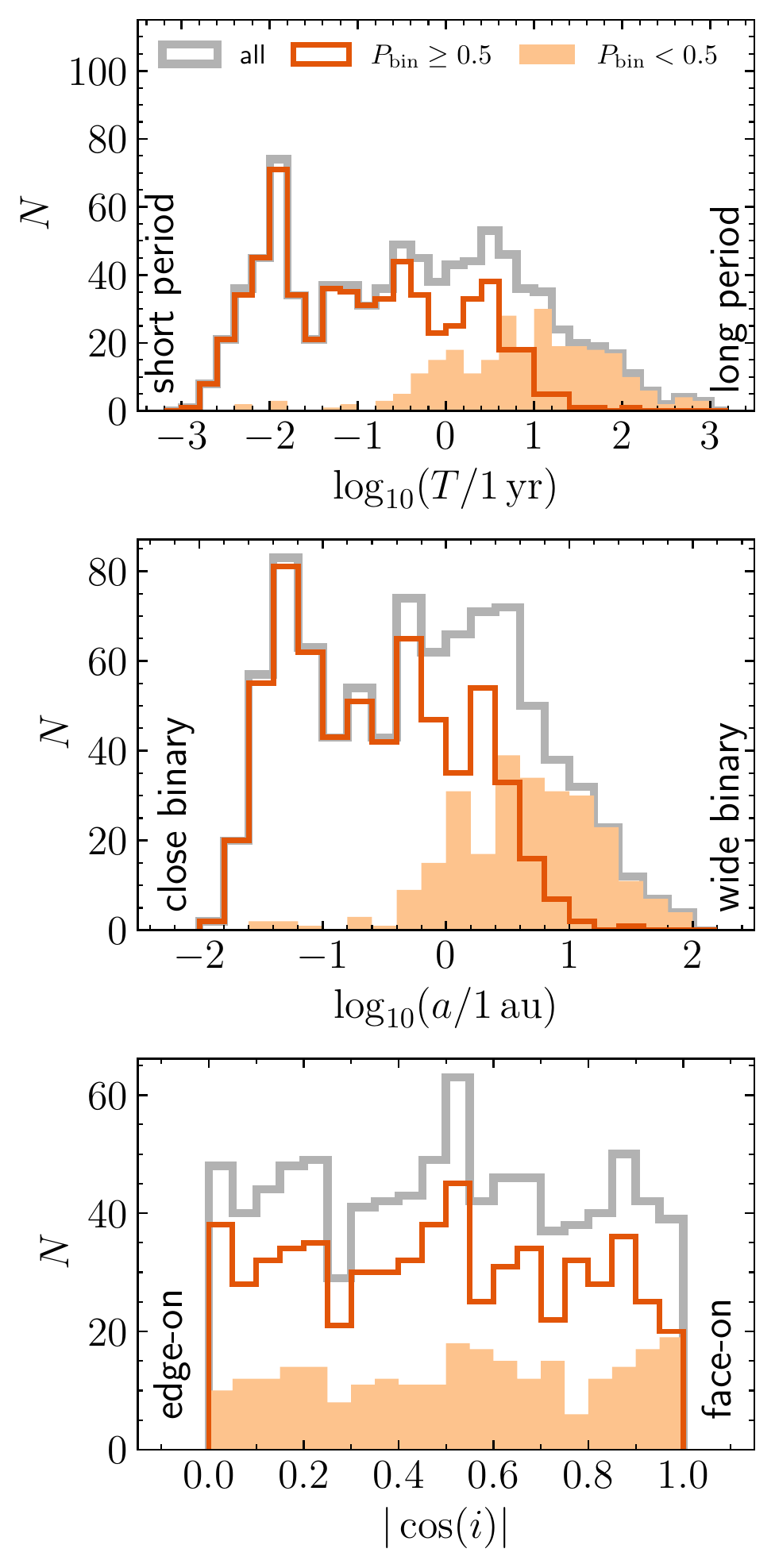}
    \caption{Period, semi-major axis and orbital plane inclination for all binaries in the same simulated GC shown in Figure \protect\ref{fig:cmd_bin}. The grey line shows all binaries in the sample, the dark-orange line shows binaries with $P_{\text{bin}}\geq0.5$ and the filled light-orange region shows binaries with $P_{\text{bin}}<0.5$. Binaries that have short periods (top panel) and that have a small semi-major axis (middle panel), i.e. hard binaries, are robustly detected with variations in RVs, as the frequency and amplitude of such variations are higher. The inclination (bottom panel), on the other hand, adds a complexity as it limits the detection for any type of binary, as binaries close to a face-on configuration ($|\cos(i)|\sim1$) are challenging to detect. In our sample, the number of binaries with $P_{\text{bin}}\geq0.5$ becomes comparable with the ones with  $P_{\text{bin}}<0.5$ for inclinations close to face-on, whereas the former dominates the sample for other inclinations.}
    \label{fig:properties_detection}
\end{figure}

Three main effects play against the detection of binaries using this approach. The first one is the luminosity ratio between the components. If both stars have similar luminosities, then the observed line-of-sight velocity given by Equation \ref{eq:obs_los} will be dominated by the centre of mass velocity rather than by their orbital motion. On the other hand, in the right panel of Figure \ref{fig:cmd_bin}, we can see that the mass ratio does not have a significant impact on the binaries detectability.

 The second effect comes from the orbital parameters of the binary, which also have a role in their detection. In Figure \ref{fig:properties_detection} we illustrate a selection of orbital parameters that have an impact on the probability of being variable in RV, and therefore on identifying binaries. We see that most of the binaries with a high probability $P_{\text{bin}}\geq0.5$ of being variable in RV have short periods and are close binaries, as shown in the top two panels of Figure \ref{fig:properties_detection}. Short period binaries orbit faster around their centre of mass, increasing the amplitude of the variability in RV. All of the binaries detected in this simulated GC have periods below $1\,\text{yr}$. The length of the semi-major axis $a$ goes in hand with the period, as binaries which survive sinking to the centre of the cluster become tighter and have shorter periods.

The third effect that affects the detectability of binaries comes from the inclination of the orbital plane. The orientation of the orbital plane is randomly chosen, therefore it is not directly related to other orbital parameters of the binaries. In the bottom panel of Figure \ref{fig:properties_detection}, all binaries in the sample have an inclination angle $i$ which follows a uniform distribution (over a sphere and, therefore, is uniform in $\cos(i)$). For most inclination angles, binaries with $P_{\text{bin}}\geq0.5$ dominate the sample. For inclination angles closer to be face-on ($|\cos{i}|\sim 1.0$) the detection becomes challenging, and  binaries with $P_{\text{bin}}<0.5$ are comparable in number to the ones with $P_{\text{bin}}\geq0.5$. This is expected as when the orbital plane becomes face-on, the line-of-sight velocity is perpendicular to the orbital motion of the binary, and no binary would be detected. 

Using this approach, we can identify binaries in each mock data set, corresponding to each simulated globular cluster.
%Figure \ref{fig:detected_binaries} shows the detected binary fraction within the half-light radius with respect to the binary fraction in each cluster at $12\,\text{Gyr}$, note that all clusters in this sample had an initial binary fraction of $f_{\text{bin}}=10\%$. 
In the following, we will discuss the effects of the binaries on the kinematics of a cluster (Section \ref{sec:kin_bin}) and their interaction with an IMBH (Section \ref{sec:imbh_bin}). 
%The difference between the two groups in Figure \ref{fig:detected_binaries} is the presence or absence of an IMBH, GCs with a low binary fraction and that follow the 1-1 line (bottom left) are the ones hosting an IMBH in the simulations. 

\section{Kinematic effects of binaries}
\label{sec:kin_bin}

As we previously discussed in \cite{aros_2020}, binaries have two main effects on the observed velocity dispersion of GCs. First, as binaries are more massive than single stars, they have, as a population, a different level of energy equipartition and hence spatial distribution. As this affects the centre of mass velocities of the binaries, both the line-of-sight velocity and the proper motions will show a lower velocity dispersion than what is expected for a cluster populated by single stars only. This effect has been studied for proper motions by \cite{bianchini_2016}. They find in GC simulations a colour bias in the velocity dispersion due to the presence of binaries, i.e. the redder edge of the main sequence (binary stars) has a lower velocity dispersion than the blue edge (single stars). However, they did not detect this in HST observations for NGC 7078, which is consistent with the low binary fraction of the cluster. In general, it is challenging to disentangle this effect, as it depends on the binary fraction and level of energy equipartition in the cluster.

The second effect is an increase in the observed velocity dispersion due to the measured line-of-sight velocity of the binaries. For binaries that have a large velocity amplitude, the orbital motion around their centre of mass will dominate the observed line-of-sight velocity, introducing a bias to larger velocity dispersion measurements. To show this effect, we construct velocity dispersion profiles from the mock data with all the stars (single+binaries) and compare it with a sample that only includes stars with a low probability of being a binary ($P_{\text{bin}}<0.3$). We exclude stars that have line-of-sight velocities outside of the central $99\%$ of the velocity distribution, centered on the cluster's mean velocity. This criterion allows us to measure the effects of binary stars without being dominated by stars with observed velocities many times the velocity dispersion of the cluster, which might not be identified as cluster members. All measured velocity dispersion values in this work follow this criterion. A stricter limit on the velocity distribution or an iterative $3\sigma$ clipping approach can reduce the binary contamination. However, our focus is the kinematic imprint of the binaries on the observed velocity dispersion and, therefore, we choose to retain most binaries for our analysis.

\begin{figure}
    \centering
    \includegraphics[width=0.9\linewidth]{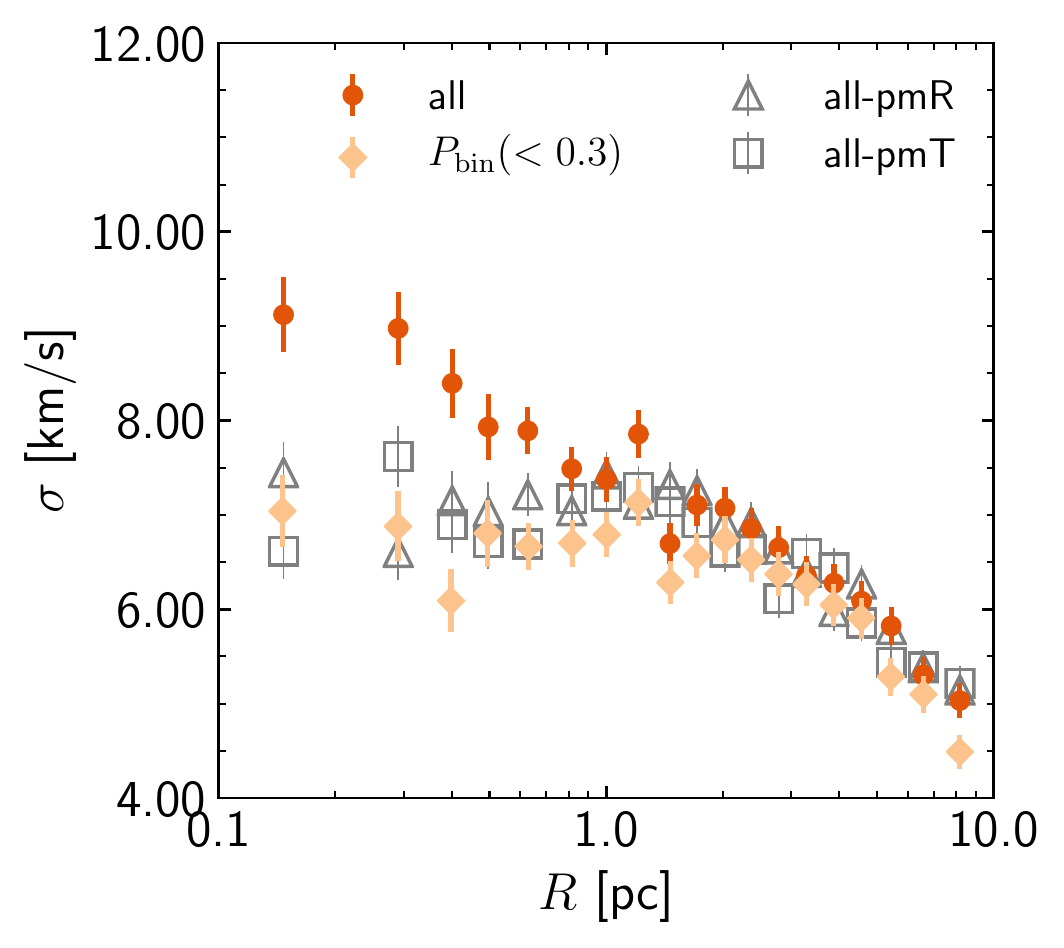}
    \caption{Line-of-sight velocity dispersion for a simulated GC with binaries (same cluster as in Figures \protect\ref{fig:cmd_bin} and \protect\ref{fig:properties_detection}). The dark-orange circles show the velocity dispersion when all stars are considered for the kinematics, this includes all binaries in the sample. Once we use the probability of being a binary assigned to each star, we can select stars that have a low probability $P_{\text{bin}}<0.3$. The light-orange diamonds show the velocity dispersion profile when most binaries are excluded. We can see that once the sample is "cleaned" from binaries, the line-of-sight velocity dispersion is consistent with the proper motion velocity dispersion (gray triangles and squares).}
    \label{fig:kin_profile}
\end{figure}

Figure \ref{fig:kin_profile} shows the observed velocity dispersion for the same cluster used as in Figure \ref{fig:cmd_bin} and \ref{fig:properties_detection}. The dark orange circles show the line-of-sight velocity dispersion for all stars in the sample (single+binaries), while the light orange diamonds show the velocity dispersion when only stars with $P_{\text{bin}}<0.3$ are selected. The difference between the whole sample and the selection with $P_{\text{bin}}<0.3$ is a consequence of the bias introduced by the orbital motion of binaries. The bias is stronger towards the cluster centre and it is a consequence of the larger binary fraction within this region. In Section \ref{sec:bin_frac_kin} we discuss further the correlation between the binary fraction and the bias in the velocity dispersion. For comparison, the radial (grey triangles) and tangential (grey squares) proper motions are also included. Both proper motions behave similarly to the sample with $P_{\text{bin}}<0.3$.

\begin{figure}
    \centering
    \includegraphics[width=0.9\linewidth]{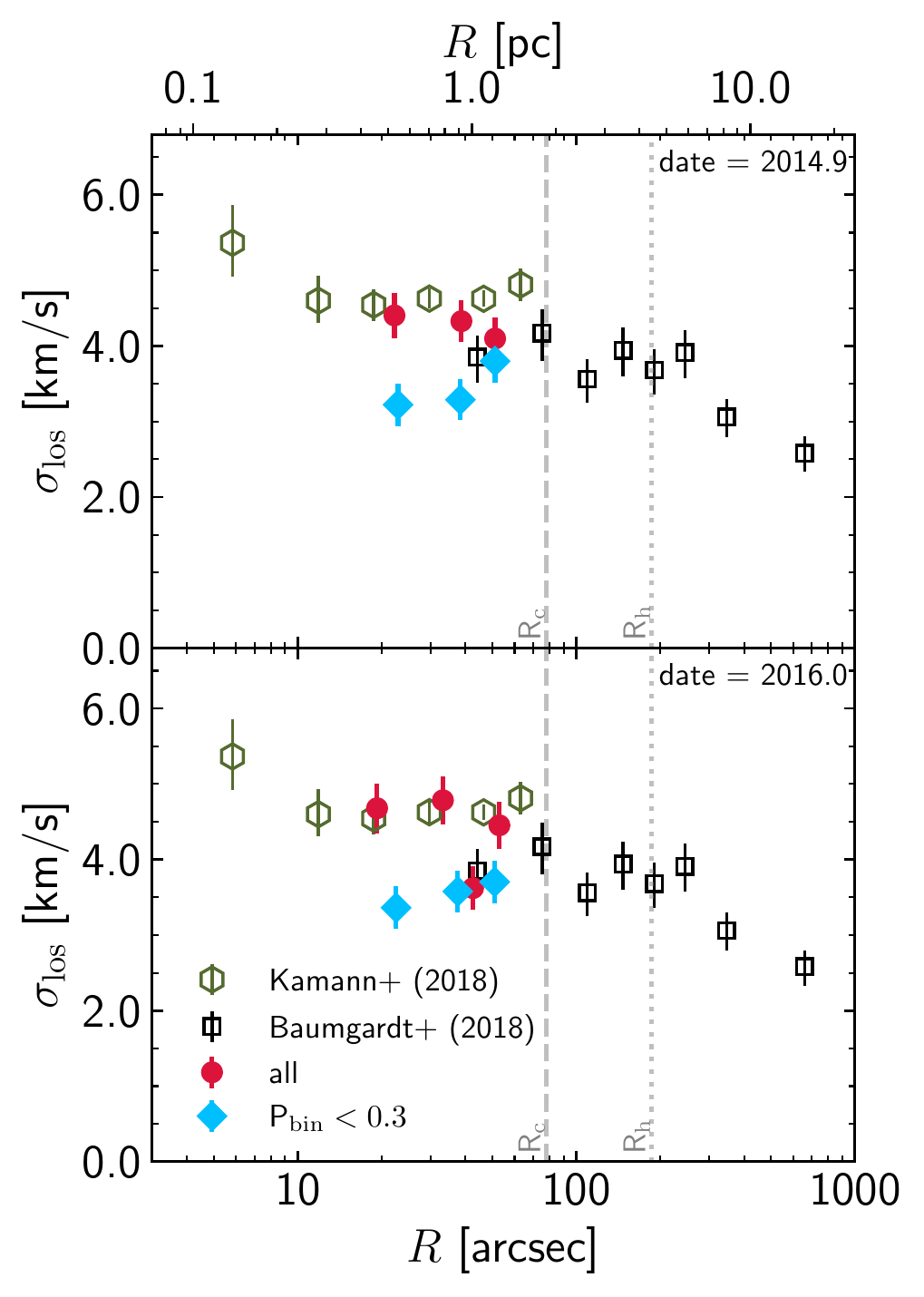}
    \caption{Line-of-sight velocity dispersion of NGC 3201. We include the data from \protect\cite{kamann_2018} (green hexagons) and \protect\cite{baumgardt_2018} (black squares), as well as the observations from \protect\cite{giesers_2019} (red circles) in two different epochs (top and bottom panel). Once we exclude stars that have a high probability of being a binary we observe a drop in the velocity dispersion similar as in Figure \protect\ref{fig:kin_profile} (blue diamonds). This is consistent for both epochs and shows the importance of multi-epoch observations to detect binaries in GCs. The grey dashed and dotted lines show the core and half-light radii of NGC 3201 from \protect\cite{harris_1996} (2010 edition).}
    \label{fig:ngc3201}
\end{figure}

While the previous example comes from mock data of a simulated globular cluster, it is also possible to observe this effect in current observations of GCs. Using multi-epoch observations of VLT/MUSE data, \cite{giesers_2019} identified binaries in NGC 3201 using the variations in radial velocity (RV). We use the available RV data from their work to analyse the kinematic effect of the binaries in their sample. Figure \ref{fig:ngc3201} shows the line-of-sight velocity dispersion for the sample in \cite{giesers_2019}, with the same approach as in Figure \ref{fig:kin_profile} we show the velocity dispersion for all stars (red circles) and for a sample of the stars with a probability of being variable in RV of $P_{\text{bin}}<0.3$ (blue diamonds). Each data point is calculated from approximately $150$ stars for the two epochs that have the largest samples of observations ($\sim 3000$ stars, upper and lower panel). Note that we only select stars with velocity errors smaller than $3\,\text{km/s}$.

From Figure \ref{fig:ngc3201}, we can see that for both epochs the sample with $P_{\text{bin}}<0.3$ indeed has a lower velocity dispersion and the difference increases towards the centre. For reference, we have also included previous measurements of the line-of-sight velocity dispersion from \cite{kamann_2018} who use VLT/MUSE data, and from \cite{baumgardt_2018} who extract velocities from archival data from ESO/VLT instruments and complement them with measurements from literature \citep{kunder_2017}. It is important to highlight that the data from \cite{giesers_2019} are an extension of the data of  \cite{kamann_2018}, which increases the number of average epochs from 7 to 12. \cite{kamann_2018} already discussed the differences between their velocity dispersion profile and the one from \cite{baumgardt_2018}, pointing to two probable sources, binaries or partial energy equipartition. For binaries, they excluded stars which were likely to be binaries based on RV variations between epochs, however, the additional epochs in \cite{giesers_2019} might have helped in detecting binaries which remained undetected by \cite{kamann_2018}, in particular short-period binaries. 

The different effects of binaries on the line-of-sight velocities and proper motions can help constraining the level of contamination caused by binaries and to estimate the binary fraction in the cluster. If, as in Figure \ref{fig:kin_profile}, a significant difference in the velocity dispersion between single epoch line-of-sight observations and proper motions is observed, then follow up line-of-sight velocity observations will be necessary to identify binaries and to clean up the kinematics. In other words, cleaning up the kinematics would be crucial for a 3D kinematic analysis of the cluster.

\section{Binary fraction and IMBHs}
\label{sec:imbh_bin}

As discussed in the previous section, binaries leave a clear signature on the observed velocity dispersion. Dynamical analysis that considers line-of-sight velocities needs to take into account their presence. While it is possible to clean the kinematic sample by means of multi-epoch observations, the identified binaries can also be utilized to help understand the dynamical state of a GC. 

\subsection{Radial distribution of binaries}
\label{sec:bin_frac_rad}

\begin{figure}
    \centering
    \includegraphics[width=1.0\linewidth]{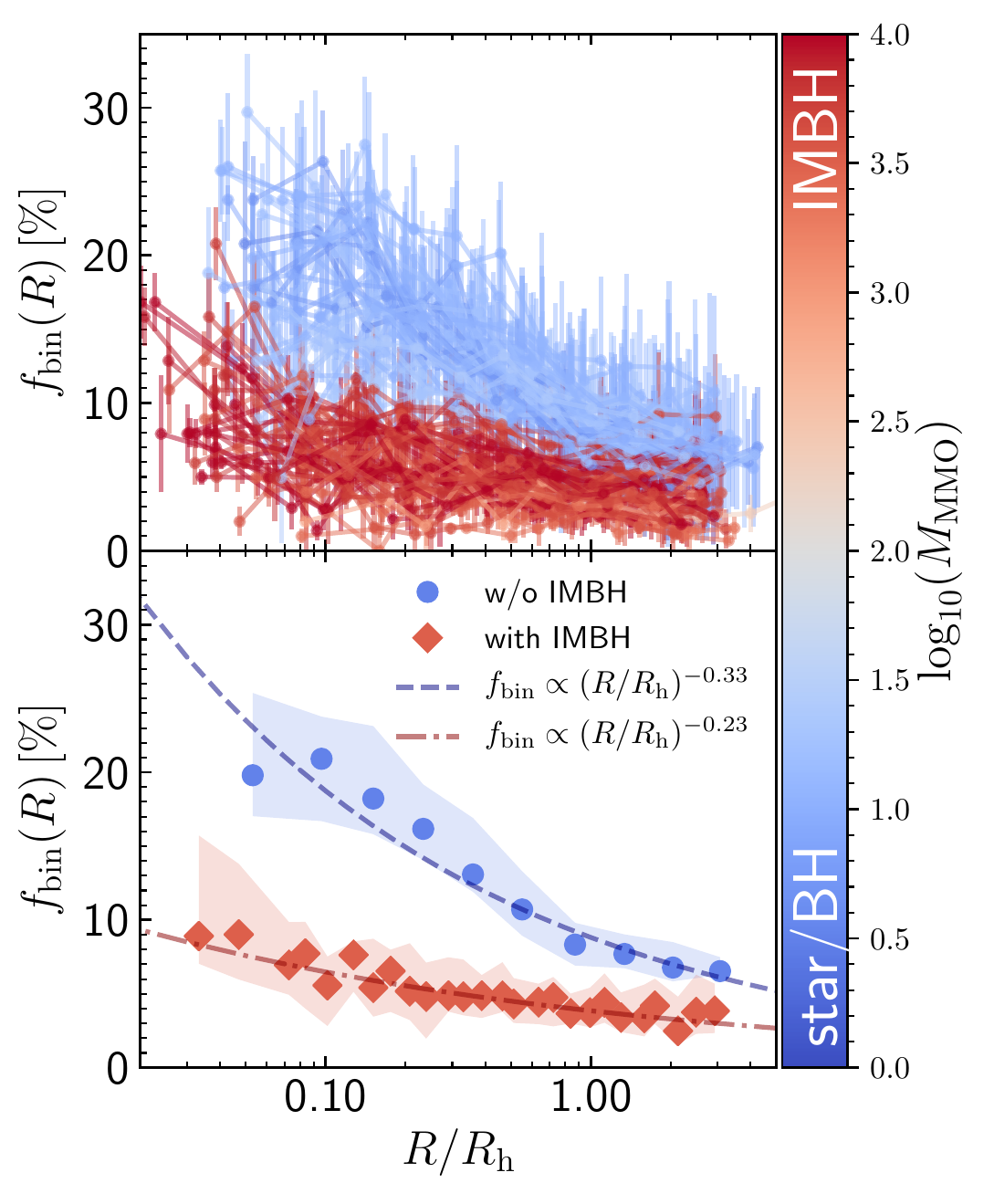}
    \caption{Radial distribution of binaries. The top panel shows the binary fraction $f_{\text{bin}}$ profile for each simulated GC in our sample, colour-coded by the mass of the most massive object ($M_{\text{MMO}}$, stars in dark blue and IMBHs in red). The symbols mark the binary fraction using $P_{\text{bin}}=0.5$ as a delimitter for binaries and non-binaries (i.e. stars with $P_{\text{bin}}\geq0.5$ are binaries, while stars with $P_{\text{bin}}<0.5$ are single stars); the error bars represent the binary fraction by separating binary and non-binary stars at $P_{\text{bin}}=0.3$ (top error bar) and $P_{\text{bin}}=0.7$ (bottom error bar). The bottom panel shows the median binary fraction profile for both types of GCs, with and without an IMBH. We fitted a power-law to the median distributions (dashed lines) and show that clusters with an IMBH have a flatter profile and a lower binary fraction than GCs without one.    }
    \label{fig:imbh_bin_prf}
\end{figure}

The binary distribution changes during the dynamical evolution of a GC, mainly due to their segregation towards the cluster’s centre and to the formation and destruction of binaries by their interaction with surrounding stars. Consequently, the cluster’s binary fraction increases towards the centre while decreasing at larger radii (see Figure \ref{fig:imbh_bin_prf}). Early \textit{N}-body simulations of GCs with binaries by \cite{heggie_2006} already showed how the half-mass radius of the binary population is indeed smaller than that of single stars. 

The presence of a central IMBH affects particularly the surviving fraction of binaries in the cluster centre, as the rate of disrupted binaries increases significantly under the presence of a central IMBH. While the increased density of stars around the IMBH mainly drives the disruption of binaries that segregate towards the cluster centre by interactions with other stars \citep[][]{trenti_2007}, close encounters of a binary with the IMBH can also break the binary producing high-velocity stars \citep[see][for example]{hills_1988,fragione_2019,subr_2019}. This implies that clusters with a central IMBH can have a reduced binary fraction towards the centre of the cluster. 

Using the detected binaries in the sample of simulated GCs, we have constructed binary fraction profiles to analyse the behaviour of GCs with and without an IMBH. The top panel of Figure \ref{fig:imbh_bin_prf} shows the binary fraction profiles of our sample of 95 simulated GCs. Each profile is colour-coded by the most massive object in the cluster (from a star of $1\,M_{\odot}$ in dark blue to an IMBH of $10^4\,M_{\odot}$ in red). In clusters without an IMBH, the most massive object is in most cases a stellar mass black hole with, on average, $\sim20\,M_{\odot}$. While all clusters started with the same fraction of $10\%$ primordial binaries, we can observe a large variety of radial distributions. However, a significant difference is the presence of an IMBH, as clusters with one clearly show a lower binary fraction at all radii as well as a flatter profile. To have more clarity in the differences of both populations of GCs (with and without IMBH), we show in the bottom panel of Figure \ref{fig:imbh_bin_prf} the median profile for each population. We find that the median distributions are well represented by a power-law ($f_{\text{bin}}\propto(R/R_{\text{h}})^{-k}$), and that clusters without an IMBH have a steeper binary fraction profile ($k=0.33\pm0.03$) compared to those with one ($k=0.23\pm0.02$).

\begin{figure}
    \centering
    \includegraphics[width=1.0\linewidth]{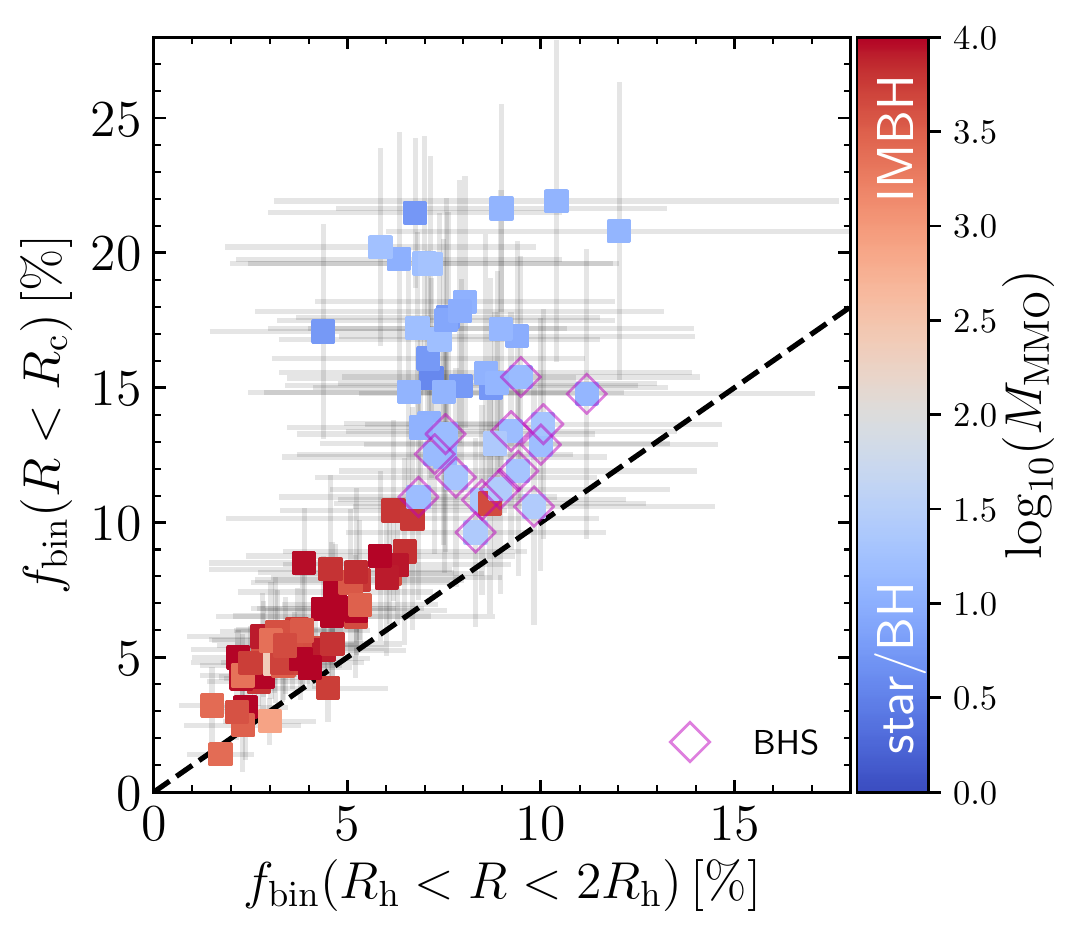}
    \caption{Binary fractions within one and two half-light radii $R_{\text{h}}$ and within the core radius $R_{\text{c}}$. Each GC is colour-coded by its most massive object. Clusters with an IMBH stay near the dashed line showing the 1-to-1 ratio between the binary fractions, i.e. less mass segregated than above the line. These clusters also populate the left side of the figure as they have systematically fewer binaries, a difference with respect to GCs with a BHS (magenta diamonds) that are near the 1-to-1 line, but have retained more binaries overall.}
    \label{fig:imbh_bin_frac}
\end{figure}

An alternative way to observe the binary fraction profile behaviour is to focus on specific regions of the GCs. From Figure \ref{fig:imbh_bin_prf},  we can see that the binary fraction at the centre and around the half-light radius matches more for clusters with an IMBH than for GCs without one. We now focus only on those two regions and measure the binary fraction within the GC’s core radius and the area within one and two half-light radii. Figure \ref{fig:imbh_bin_frac} shows the binary fractions in these regions for all clusters in our sample, plotted against each other. Once again this is colour-coded by the most massive object in the GC. We can see that GCs with a central IMBH (in red) have a lower binary fraction and populate the figure’s bottom-left region; furthermore, most of these clusters are close to the 1-to-1 relation (dashed line). Clusters without an IMBH have a higher binary fraction, particularly in the core, and move away from the 1-to-1 line, which shows the expected effects and consequences of mass segregation in the binary population. A group of GCs without an IMBH is also located close to the 1-to-1 line. These clusters have retained many stellar-mass black holes instead of an IMBH, and to identify them, we have marked with a magenta diamond all GC with more than 50 stellar remnant black holes (i.e. a black hole system, BHS) within the half-mass radius. While they are closer to the 1-to-1 line with respect to clusters without a BHS, they still have a higher binary fraction within the core than clusters with an IMBH. On the other hand, they have a lower binary fraction than GCs without so many central black holes or an IMBH. Whereas the BHS works as an energy source for the cluster to halt mass segregation, the density within the core is not high enough to trigger an efficient disruption of binaries \citep[see e.g.][]{mackey_2008,sipel_2013,breen_2013,morscher_2015,weatherford_2018}. 

\begin{figure}
    \centering
    \includegraphics[width=1.0\linewidth]{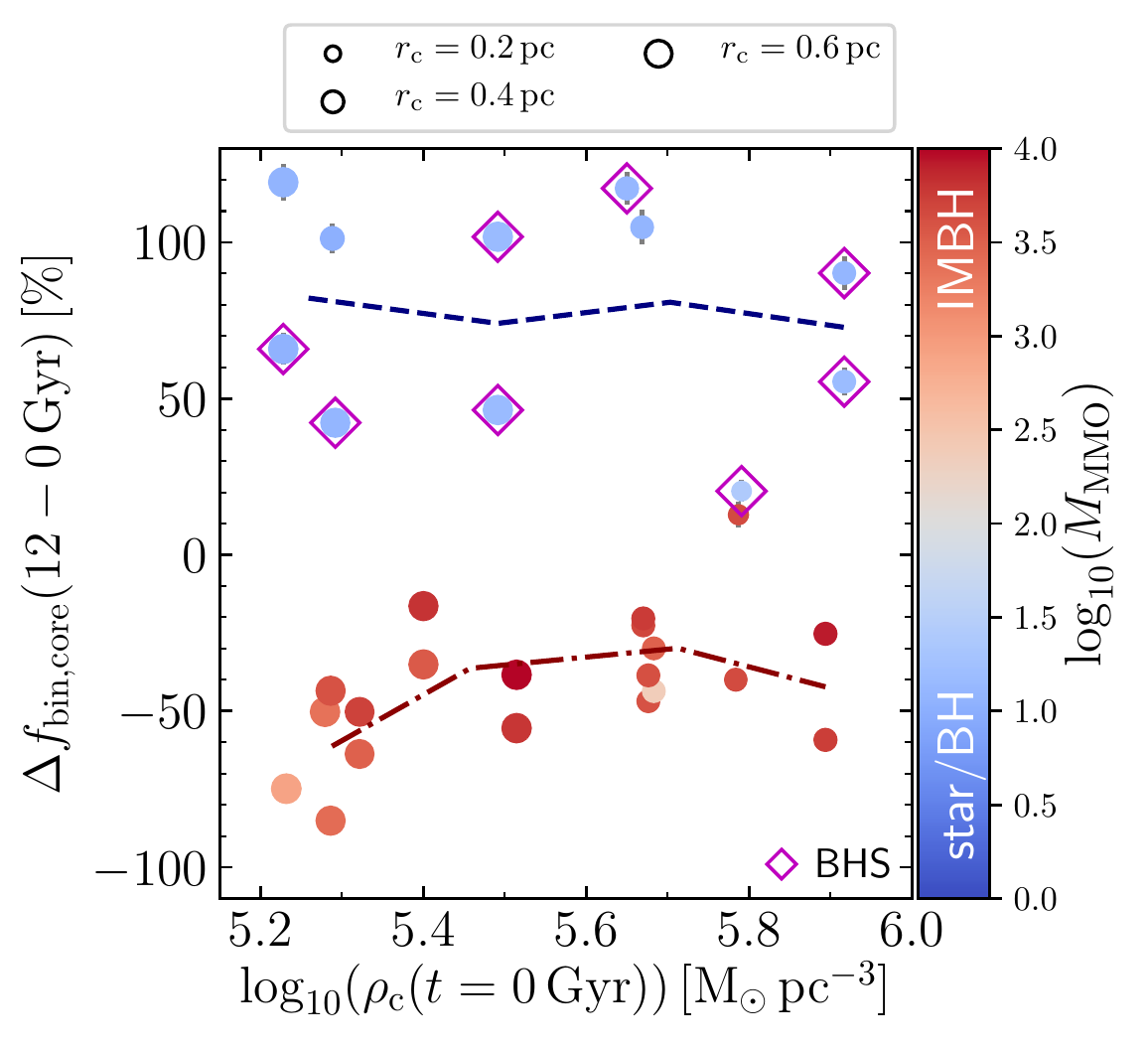}
    \caption{Percentage difference between the initial binary fraction and the binary fraction at $12\,\text{Gyr}$ within the core radius as given by Equation \protect\ref{eq:delta_fbin}. We selected a subsample of 32 simulated clusters with comparable initial central density. The binary fraction at $12\,\text{Gyr}$ corresponds to the observed binary fraction within the core radius as in Figure \protect\ref{fig:imbh_bin_frac}. All GCs are colour coded by their most massive object.  The sizes of the symbols indicate the initial core radius of the GCs, and the magenta diamonds highlight the GCs with a BHS.  The red dot-dashed and blue dashed lines represent the mean percentage difference for the population of simulated GCs with and without an IMBH, respectively. The binary fraction within the core radius of GCs with an IMBH halves with respect to their initial conditions. For those without an IMBH, we observe a percentage increase of $70\%$ with respect to the initial value.}
    \label{fig:lost_binaries}
\end{figure}

The central density plays an active role in the dynamical interaction and disruption of binary systems. Therefore, to better isolate the effect of an IMBH on the binary fraction, we select a subsample with comparable initial central density. In our sample, GCs with initial central densities below $\rho_{\text{c}}\sim10^5\,M_{\odot}/\text{pc}^3$ do not form an IMBH within $12\,\text{ Gyr}$, while most if not all clusters with initial central densities larger than $\rho_{\text{c}}\sim10^6\,M_{\odot}/\text{pc}^3$ form an IMBH. In the region between these two limits we have 32 GCs which can either form an IMBH or not. Figure \ref{fig:lost_binaries} shows the percentage difference between the initial binary fraction and the binary fraction at $12\,\text{Gyr}$, both calculated within the core radius and defined as:
\begin{equation}
    \Delta f_{\text{bin,core}}(12-0\,\text{Gyr}) = \frac{f_{\text{bin,core}}(12) - f_{\text{bin,core}}(0)}{f_{\text{bin,core}}(0)}\,,\label{eq:delta_fbin}
\end{equation}
where $f_{\text{bin,core}}(t)$ is the binary fraction within the core radius at a given time in $\text{Gyr}$. GCs that form an IMBH have mainly negative percentage differences, and on average, they have lost $\sim50\%$ of the initial binary fraction. On the other hand, GCs that do not form a central IMBH have their core binary fraction increased by $\sim70\%$ with respect to the initial binary fraction. In this range of initial central densities, the IMBH forms via the “SLOW” scenario late in the evolution of the GC (see Section \ref{sec:simulations} and references therein), preceded by an increased central density due core-collapse. For clusters that follow the “SLOW” scenario, we observe that the binary fraction within the core radius increases steadily until the central density is high enough to trigger the formation of an IMBH. During this process, the binary fraction decreases rapidly along with an increase in binary-single and binary-binary interactions. On the other hand, for GCs in Figure \ref{fig:lost_binaries} that do not form an IMBH, the presence of stellar-mass black holes helps delaying the core-collapse and quenches the central density rise, preventing the formation of an IMBH and the depletion of binaries.

\begin{figure}
    \centering
    \includegraphics[width=0.9\linewidth]{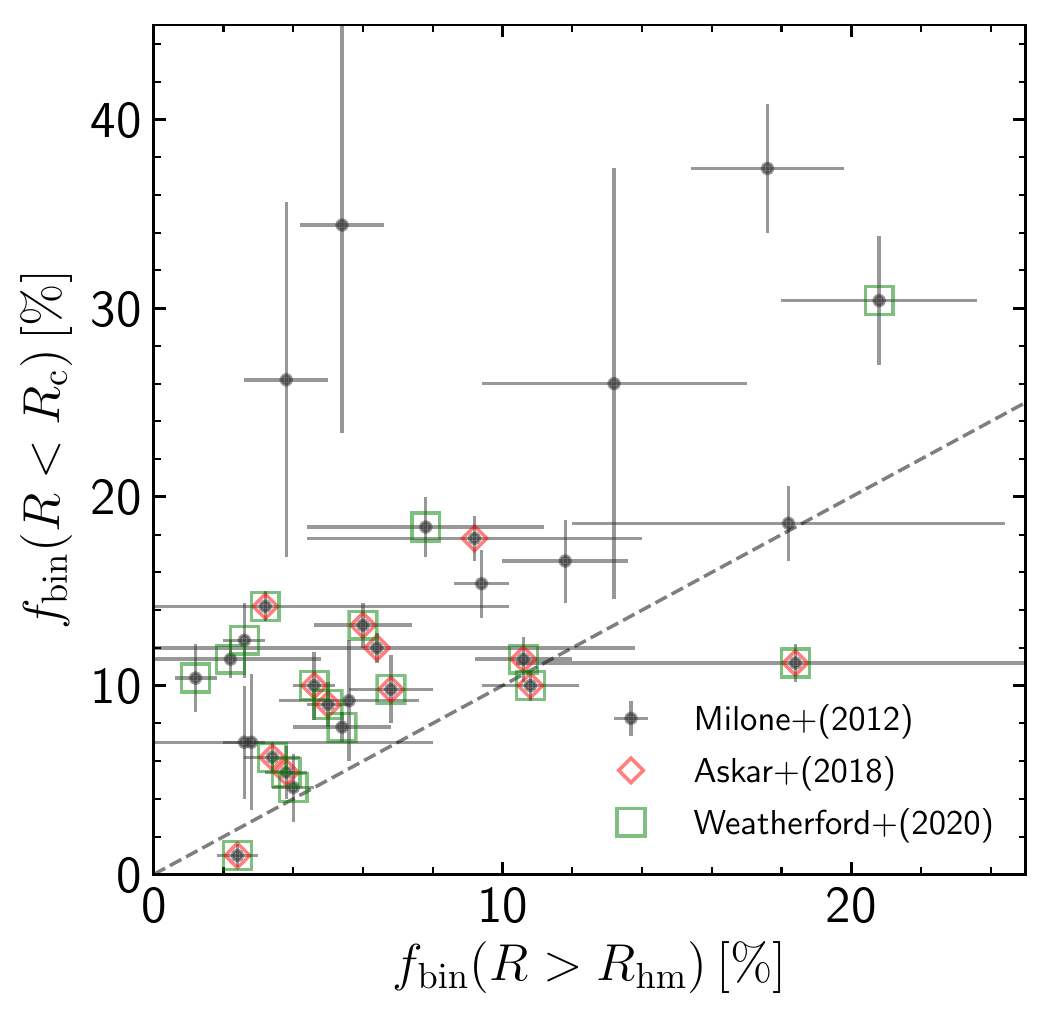}
    \caption{Binary fraction within the core radius $R_{\text{c}}$ and outside the half-mass radius $R_{\text{hm}}$ for Galactic globular clusters. The binary fractions measured by \protect\cite{milone_2012} show the same behaviour as Figure \protect\ref{fig:imbh_bin_frac}, almost all GCs are above the 1-to-1 dashed line, following the expected effect of mass segregation with the exception of NGC 288, see Section 4.1 for more details. We have marked GC candidates that may be hosting a BHS as given by \protect\cite{askar_2018b} (red diamond) and \protect\cite{weatherford_2020} (green squares). Most of the candidate GCs are located near the 1-to-1 line, in a similar way as in Figure \protect\ref{fig:imbh_bin_frac}.}
    \label{fig:bin_frac_milone}
\end{figure}

The advantage of using integrated quantities over different regions of the cluster is that this technique can be applied to explore the behaviour of observed GCs. \cite{milone_2012} surveyed the binary fraction of GCs using HST photometry along the main sequence. While there are intrinsic differences in the detected binaries from their method and the one described here, this is a first-order approach to show that it is possible to test our results with photometric data. Figure \ref{fig:bin_frac_milone} shows the binary fraction within the core radius and the region outside the half-mass radius, as given by \cite{milone_2012}. We see that almost all clusters are near or above the 1-to-1 relation (dashed line). Only NGC 288 falls clearly below the 1-to-1 line. While the error bars are big enough to include the 1-to-1 line, it is essential to keep in mind that this cluster has tidal tails \citep[see][for the latest discussion using Gaia data]{kaderali_2019,sollima_2020} and could have undergone a more complex dynamical evolution than other clusters in the sample. We compared the clusters in \cite{milone_2012} with two studies that estimate the number of remaining stellar mass black holes: \cite{askar_2018b} (red diamonds) use properties such as the core radius and the central density to estimate the number of retained BHs, whereas \cite{weatherford_2020} (green squares) use the mass segregation of the clusters as an indicator for the remaining BHs.  Although both studies estimate different numbers of BHs (see Table \ref{tab:bin_frac} in the Appendix), the clusters with retained BHs are near the 1-to-1 line. Most Galactic clusters with candidate IMBHs do not have a measured binary fraction within the core and have been excluded from Figure \ref{fig:bin_frac_milone}.  

\subsection{Kinematic effects due to binaries}
\label{sec:bin_frac_kin}
We discussed how the binary fractions can point to GCs that could host an IMBH or a BHS. However, the observed kinematics are another piece of information that could also show the binaries interaction with a central IMBH. As we discussed in Section \ref{sec:kin_bin}, binaries affect the line-of-sight and proper motion velocity dispersions differently. Figure \ref{fig:kin_profile} shows that the difference between line-of-sight and proper motion velocity dispersions increases significantly towards the centre, which goes in hand with the increase in the binary fraction in the cluster centre. To analyse this effect, we have measured the velocity dispersion within the core radius of each GC, following the same procedure as in Section \ref{sec:kin_bin}. We define a percentage difference for the velocity dispersion given by $(\sigma_{\text{los}}-\sigma_{\text{ref}})/\sigma_{\text{ref}}$, where $\sigma_{\text{ref}}$ is a reference velocity dispersion, which can be either from proper motions or a clean sample of line-of-sight velocities. 

\begin{figure}
    \centering
    \includegraphics[width=1.0\linewidth]{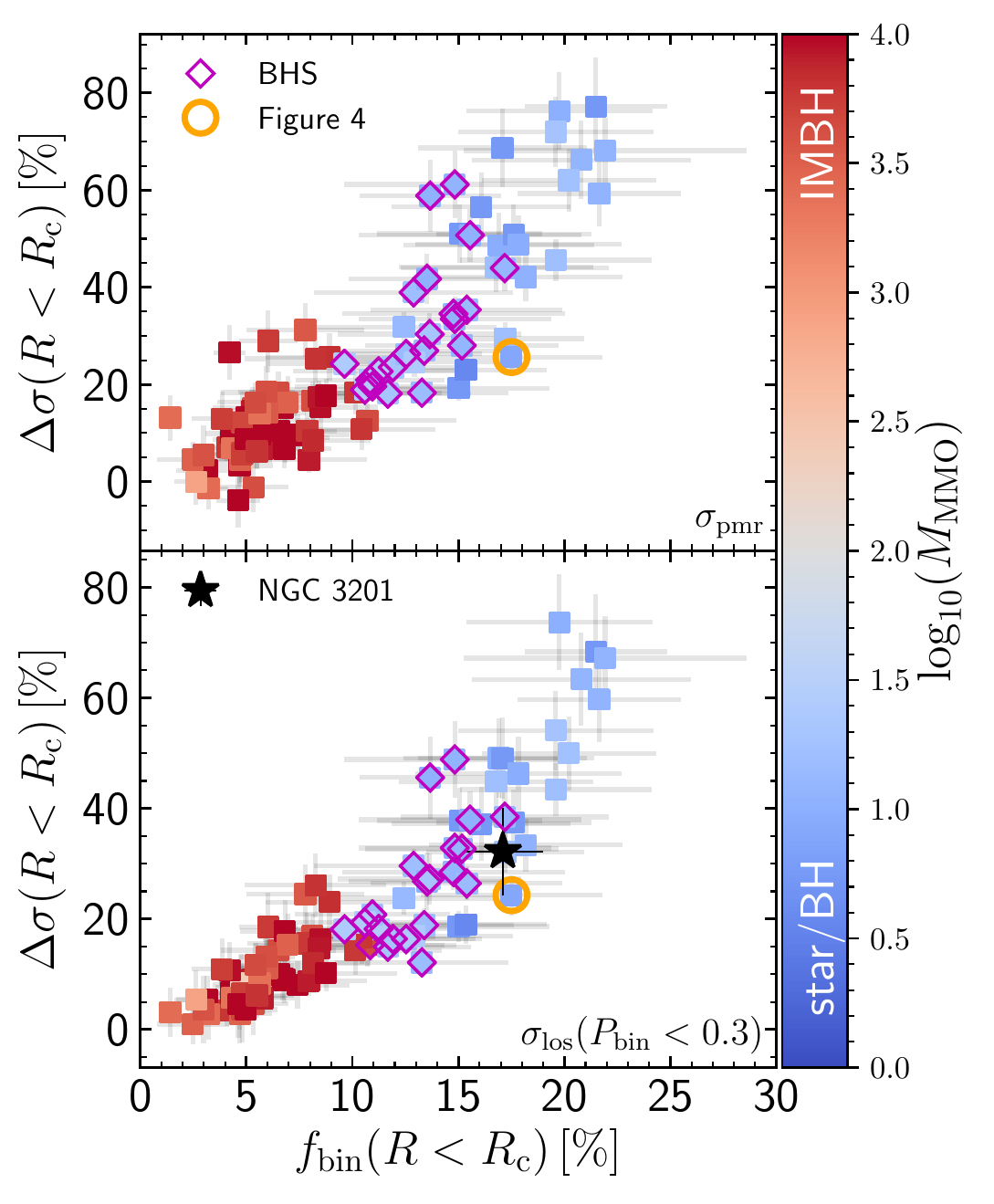}
    \caption{Difference in velocity dispersion and binary fraction within the core radius. %In Section \ref{sec:kin_bin} we show the kinematic effect of binaries in the velocity dispersion when not excluded from the kinematic sample (Figure \protect\ref{fig:kin_profile}). 
    By using the differences in velocity dispersion between the full line-of-sight contaminated sample and the proper motions (top panel) or the clean sample with $P_{\text{bin}}<0.3$ (bottom panel). All clusters are again colour-coded by their most massive object. In both panels, as the binary fraction within the core increases also does the difference in velocity dispersion, for some clusters the measured velocity dispersion with binaries overestimates up to $70\%$ the core velocity dispersion. Using proper motions as a reference adds an additional scatter due to variations in the central velocity anisotropy, but when available could provide a first constraint on the velocity difference (in particular when multi-epoch observations are not available). A more self-consistent approach, once multi-epoch observations are available, would be to compare the contaminated velocity dispersion with the clean one as in the bottom panel. This approach is not affected by velocity anisotropy and can be used right away for current and upcoming MUSE data, such as for NGC 3201 \protect\citep{giesers_2019} marked as a star. 
    }
    \label{fig:imbh_bin_kin}
\end{figure}

The top panel of Figure \ref{fig:imbh_bin_kin} shows the case when we compare the single epoch line-of-sight velocity dispersion (i.e. including the effect of binaries) with the radial proper motion. We observe that as the binary fraction increases, the difference between the velocity dispersions becomes larger. As discussed before, clusters with an IMBH have a lower binary fraction, and now we can see that they also have a smaller difference between the line-of-sight and proper motion velocity dispersions. GCs with a BHS (magenta diamonds) populate the region between the clusters with an IMBH and those without one. Obtaining comparable observations for line-of-sight velocities and proper motions is not straightforward, as not many cover the same region of the cluster or even the same stellar mass range. Due to the GC’s evolution towards energy equipartition, it is crucial to sample the same stellar mass range for both velocities. A way to move around this issue is to only focus on the line-of-sight velocities, given multi-epoch observations to identify binaries. The difference between the line-of-sight velocity dispersion that includes all binaries and the cleaned sample will play the same role as the proper motion data. The bottom panel of Figure \ref{fig:imbh_bin_kin} shows the case when the reference velocity dispersion for $\Delta\sigma$ is the line-of-sight velocity dispersion after cleaning the binaries as described in Section \ref{sec:kin_bin}. The GCs’ behaviour is similar to the previous case, but it only depends on one observational data type, which is the RVs. In this case, we can now also include the kinematic data for NGC 3201 from \cite{giesers_2019}. We measured the binary fraction and velocity dispersion difference within the core radius from the same stellar sample as in Figure \ref{fig:ngc3201}, finding that the binary fraction is $f_{\text{bin}}(R<R_{\text{c}})=17.1\pm1.9\%$ while the velocity dispersion difference is $\Delta\sigma(R<R_{\text{c}})=32.2\pm7.8\%$. As shown in the bottom panel of Figure \ref{fig:imbh_bin_kin}, this cluster follows a relation in line with the simulated GCs. This approach could be more accessible as more GCs get observed in the same way as done by \cite{giesers_2019}.

The most extreme cases in Figure \ref{fig:imbh_bin_kin} show an overestimation of $\sim70\%$ of the observed line-of-sight velocity dispersion due to the effects of binaries. However, our analysis assumes a strong binary contamination, where we only exclude stars outside the central $99\%$ of the observed velocity distribution (equivalent to a single $3\sigma$ clipping). A stricter limit can exclude most binaries, and the bias shown in Figure \ref{fig:imbh_bin_kin} will be milder. We estimated the clusters’ velocity dispersion values using an iterative $3\sigma$ clipping approach, significantly reducing the binary contamination. While the trend was still similar to the one shown in the bottom panel of Figure \ref{fig:imbh_bin_kin}, the most extreme cases went down from $\sim70\%$ to a $\sim20\%$ difference in velocity dispersion.

\section{Summary}
\label{sec:summary}
In this work, we explored the dynamical effects of a central IMBH on the binary population of GCs. We used a sample of simulated GCs from the \textsc{MOCCA}-Survey Database I \citep{askar_2017a} and applied the method proposed by \cite{giesers_2019} to identify binaries through multi-epoch radial velocity (RV) observations. We have produced mock observations for each cluster considering velocity errors in RVs and proper motions; for these mock observations, we follow the radial velocities during many epochs, considering long-term observations spanning over months as well as multiple observations during a single night. This approach allows us to identify binaries in the simulated GCs in the same manner as done for real observations.

Two-body relaxation drives the dynamical evolution of GCs; a consequence of this process is mass segregation. Binaries are on average more massive than single stars and segregate quickly towards the centre of the cluster. This phenomenon changes the distribution of binaries and the radial binary fraction. Early \textit{N}-body simulations of GCs with primordial binaries show binaries migrate towards the cluster core while becoming harder \citep{heggie_2006}. Observations of main-sequence star binaries also show this trend \citep{sollima_2007,milone_2012,ji_2015}, as the binary fraction is smaller in regions outside the core radius. The higher fraction of binaries within the cluster core affects the observed line-of-sight kinematics, as the relative velocity of hard binaries dominates over their centre of mass velocity, increasing the observed velocity dispersion along the line-of-sight (see Figure \ref{fig:kin_profile} and discussion therein). While this effect does not affect proper motions, to take full advantage of proper motions and radial velocities, the identification of binaries is crucial. Whereas the effects on the dynamical modelling are beyond the scope of this work, we expect that due to the increase in the observed line-of-sight velocity dispersion, dynamical models will overestimate the cluster mass (larger mass-to-light ratio) and (or) the mass of a possible IMBH.

An IMBH in the cluster centre acts as an energy source, extending the cluster dynamical life. It halts mass segregation, and therefore, fewer binaries move to the cluster centre, which reduces the binary fraction in the cluster core. Furthermore, the binaries that do segregate encounter a dense environment where interactions with other stars can disrupt them, due to the high central density in GCs that form an IMBH. If a binary manages to get close enough to the IMBH, this could also break the binary and one of the components might be ejected at high velocities \citep{hills_1988,fragione_2019,subr_2019}. Our analysis shows that GCs with a central IMBH have a significantly lower binary fraction than clusters without one (see Figure \ref{fig:imbh_bin_prf}). Furthermore, for GCs with comparable initial central density, we observe that those which form an IMBH lose $\sim50\%$ of their initial binary fraction within the cluster core, while the ones which do not form an IMBH increase their binary fraction within the cluster core by $\sim70\%$. 

We find that clusters that host a black hole system behave in a similar way as those with an IMBH and also show a flat binary fraction profile. However, as binary disruption is less effective than in clusters with a central IMBH and the presence of stellar-mass black holes in the cluster's core effectively halts the increase in central density, GCs with a BHS have on average higher binary fractions in their cores. This is in agreement with previous simulations of GCs with BHS \citep[see, e.g., ][]{mackey_2008,morscher_2015}. The comparison of integrated quantities, such as the binary fraction within the central and outer regions of the GCs, serves as an indication for which cluster may host an IMBH or a BHS (see Figure \ref{fig:imbh_bin_frac}). We find a handful of Galactic GCs are promising candidates, as they have a similar binary fraction within the cluster’s core and outside of the half-mass radius (see Figure \ref{fig:bin_frac_milone}). A list of the names and properties of these clusters can be found in Appendix \ref{secA:table} (see Table \ref{tab:bin_frac}).

By also taking into account the kinematic data, we can have a similar picture of the dynamical effect of an IMBH on the binary population of a cluster. Due to the lower binary fractions in clusters with a central IMBH, and in particular within the cluster core, we expect that the difference between line-of-sight and proper motion velocity dispersion will be small. We show that this is the case in Figure \ref{fig:imbh_bin_kin}. However, comparing line-of-sight velocities and proper motions is not trivial. Different effects such as the intrinsic velocity anisotropy, the cluster regions covered by different observations, the magnitude range that translates to different masses for different observations, and precise distances could add a systematic bias in the observed difference between line-of-sight and proper motion velocity dispersion. 

A solution to these limitations will be to use only line-of-sight velocities once multi-epoch observations will be available as shown in the bottom panel of Figure \ref{fig:imbh_bin_kin}. This approach will not only help as a method to identify systems that may host an IMBH or a BHS, but by knowing the binary fraction it will be possible to estimate the increase in velocity dispersion due to the effect of binaries; this is crucial for dynamical modelling that uses line-of-sight velocity as a tracer. 

We have focused only on binaries brighter than one magnitude below the main-sequence turn-off ($m_{V}<18\,\text{mag}$ at $5\,\text{kpc}$). The different mass range and binary detection biases limit the comparison with photometric studies \cite[such as][]{sollima_2007,milone_2012,ji_2015} that typically focus on the main-sequence. Extending our analysis to other detection techniques for binaries will provide a better link between the estimated binary fraction and kinematics effects of binaries. To do so, we first need to consider multiple populations in the simulations \citep[][]{hong_2015} as they play a relevant role in the photometric binary detection \citep{milone_2020}. At the same time, it is necessary to check the different biases in the detection methods: radial velocity methods mainly detect binaries with short periods, whereas photometric methods depend on the mass ratio of binaries \citep{milone_2012}. 

We analysed simulations with initial binary fractions of $f_{\text{bin}}=10\%$. These clusters have total binary fractions of $f_{\text{bin}}\sim7\%$ at $12\,\text{Gyr}$ (or $f_{\text{bin}}\sim16\%$ within the half-light radius). Whereas our initial binary fraction is consistent with observations, clusters that have higher binary fractions exist (see Figure \ref{fig:bin_frac_milone}). To compare with these observations, we need to study simulations with different initial binary fractions. See Appendix \ref{secA:fbin_comparison} for a comparison with a sample of clusters with initial $f_{\text{bin}}=5\%$, where we observe the same overall trends.  

Here we presented the properties of the whole detected binary population. The next step will be to study how the relations described here connect with specific types of binaries or binary products. The distribution of blue-straggler stars traces the dynamical evolution of GCs \cite{ferraro_2012,ferraro_2018}. By analysing the distribution of blue-straggler stars and their relation with the binary fraction, it will be possible to study another tracer for the dynamical effects of a central IMBH. Similarly, the distribution of X-ray binaries could help separate clusters with a central IMBH from those with a BHS.

\section*{Acknowledgements}

We thank the anonymous referee for their constructive comments and suggestions, which helped us improve this manuscript. We thank the \textsc{MOCCA}-Survey collaboration for making their data available to us and answering all our questions. FIA also thanks Sebastian Kamann for constructive discussions regarding the MUSE data, Alice Zocchi for her helpful suggestions regarding the manuscript and Nadine Neumayer for helpful discussions and ideas. FIA and GvdV acknowledge funding from the European Research Council (ERC) under the European Union's Horizon 2020 research and innovation programme under grant agreement No 724857 (Consolidator Grant ArcheoDyn). ACS is supported by the Deutsche Forchungsgemeinschaft (DFG, German Research Fundation) -- Project-ID 138713538 -- SFB 881 ("The Milky Way System", subproject A08). AMB is supported by the Swedish Research Council (grant 2017-04217) and acknowledges funding from the European Union’s Horizon 2020 research and innovation programme under the Marie Skłodowska-Curie grant agreement No 895174. AA is supported by the Swedish Research Council (grant 2017-04217) and also acknowledges partial support from the Royal Physiographic Society of Lund and the Walter Gyllenberg Foundation for the research grant: "Evolution of Binaries containing Massive Stars".

This research made use of the \textsc{NUMPY} package \citep{van_der_walt_2011}, while all figures were made using \textsc{MATPLOTLIB} \citep{hunter_2007}.

%%%%%%%%%%%%%%%%%%%%%%%%%%%%%%%%%%%%%%%%%%%%%%%%%%
\section*{Data Availability}

The simulated GCs data underlying this article were provided by the \textsc{MOCCA} group\footnote{\url{https://moccacode.net/}} by permission. The data will be shared upon request with permission of the \textsc{MOCCA} group. The observational data is available within their original references. 
 
%The inclusion of a Data Availability Statement is a requirement for articles published in MNRAS. Data Availability Statements provide a standardised format for readers to understand the availability of data underlying the research results described in the article. The statement may refer to original data generated in the course of the study or to third-party data analysed in the article. The statement should describe and provide means of access, where possible, by linking to the data or providing the required accession numbers for the relevant databases or DOIs.

%%%%%%%%%%%%%%%%%%%%%%%%%%%%%%%%%%%%%%%%%%%%%%%%%%



\bibliographystyle{mnras}
\bibliography{IMBH2} % if your bibtex file is called example.bib

\begin{thebibliography}{}
\makeatletter
\relax
\def\mn@urlcharsother{\let\do\@makeother \do\$\do\&\do\#\do\^\do\_\do\%\do\~}
\def\mn@doi{\begingroup\mn@urlcharsother \@ifnextchar [ {\mn@doi@}
  {\mn@doi@[]}}
\def\mn@doi@[#1]#2{\def\@tempa{#1}\ifx\@tempa\@empty \href
  {http://dx.doi.org/#2} {doi:#2}\else \href {http://dx.doi.org/#2} {#1}\fi
  \endgroup}
\def\mn@eprint#1#2{\mn@eprint@#1:#2::\@nil}
\def\mn@eprint@arXiv#1{\href {http://arxiv.org/abs/#1} {{\tt arXiv:#1}}}
\def\mn@eprint@dblp#1{\href {http://dblp.uni-trier.de/rec/bibtex/#1.xml}
  {dblp:#1}}
\def\mn@eprint@#1:#2:#3:#4\@nil{\def\@tempa {#1}\def\@tempb {#2}\def\@tempc
  {#3}\ifx \@tempc \@empty \let \@tempc \@tempb \let \@tempb \@tempa \fi \ifx
  \@tempb \@empty \def\@tempb {arXiv}\fi \@ifundefined
  {mn@eprint@\@tempb}{\@tempb:\@tempc}{\expandafter \expandafter \csname
  mn@eprint@\@tempb\endcsname \expandafter{\@tempc}}}

\bibitem[\protect\citeauthoryear{{Arca Sedda}, {Askar}  \& {Giersz}}{{Arca
  Sedda} et~al.}{2019}]{arca_sedda_2019}
{Arca Sedda} M.,  {Askar} A.,   {Giersz} M.,  2019, arXiv e-prints, \href
  {https://ui.adsabs.harvard.edu/abs/2019arXiv190500902A} {p. arXiv:1905.00902}

\bibitem[\protect\citeauthoryear{{Aros}, {Sippel}, {Mastrobuono-Battisti},
  {Askar}, {Bianchini}  \& {van de Ven}}{{Aros} et~al.}{2020}]{aros_2020}
{Aros} F.~I.,  {Sippel} A.~C.,  {Mastrobuono-Battisti} A.,  {Askar} A.,
  {Bianchini} P.,   {van de Ven} G.,  2020, \mn@doi [\mnras]
  {10.1093/mnras/staa2821}, \href
  {https://ui.adsabs.harvard.edu/abs/2020MNRAS.499.4646A} {499, 4646}

\bibitem[\protect\citeauthoryear{{Askar}, {Szkudlarek}, {Gondek-Rosi{\'n}ska},
  {Giersz}  \& {Bulik}}{{Askar} et~al.}{2017}]{askar_2017a}
{Askar} A.,  {Szkudlarek} M.,  {Gondek-Rosi{\'n}ska} D.,  {Giersz} M.,
  {Bulik} T.,  2017, \mn@doi [\mnras] {10.1093/mnrasl/slw177}, \href
  {https://ui.adsabs.harvard.edu/abs/2017MNRAS.464L..36A} {464, L36}

\bibitem[\protect\citeauthoryear{{Askar}, {Giersz}, {Pych}  \& {Dalessand
  ro}}{{Askar} et~al.}{2018a}]{askar_2018}
{Askar} A.,  {Giersz} M.,  {Pych} W.,   {Dalessand ro} E.,  2018a, \mn@doi
  [\mnras] {10.1093/mnras/sty101}, \href
  {https://ui.adsabs.harvard.edu/abs/2018MNRAS.475.4170A} {475, 4170}

\bibitem[\protect\citeauthoryear{{Askar}, {Arca Sedda}  \& {Giersz}}{{Askar}
  et~al.}{2018b}]{askar_2018b}
{Askar} A.,  {Arca Sedda} M.,   {Giersz} M.,  2018b, \mn@doi [\mnras]
  {10.1093/mnras/sty1186}, \href
  {https://ui.adsabs.harvard.edu/abs/2018MNRAS.478.1844A} {478, 1844}

\bibitem[\protect\citeauthoryear{{Baumgardt} \& {Hilker}}{{Baumgardt} \&
  {Hilker}}{2018}]{baumgardt_2018}
{Baumgardt} H.,  {Hilker} M.,  2018, \mn@doi [\mnras] {10.1093/mnras/sty1057},
  \href {https://ui.adsabs.harvard.edu/abs/2018MNRAS.478.1520B} {478, 1520}

\bibitem[\protect\citeauthoryear{{Baumgardt}, {Makino}  \&
  {Ebisuzaki}}{{Baumgardt} et~al.}{2004}]{baumgardt_2004}
{Baumgardt} H.,  {Makino} J.,   {Ebisuzaki} T.,  2004, \mn@doi [\apj]
  {10.1086/423299}, \href
  {https://ui.adsabs.harvard.edu/abs/2004ApJ...613.1143B} {613, 1143}

\bibitem[\protect\citeauthoryear{{Baumgardt} et~al.,}{{Baumgardt}
  et~al.}{2019}]{baumgardt_2019}
{Baumgardt} H.,  et~al., 2019, \mn@doi [\mnras] {10.1093/mnras/stz2060}, \href
  {https://ui.adsabs.harvard.edu/abs/2019MNRAS.488.5340B} {488, 5340}

\bibitem[\protect\citeauthoryear{{Belczynski}, {Kalogera}  \&
  {Bulik}}{{Belczynski} et~al.}{2002}]{belczynski_2002}
{Belczynski} K.,  {Kalogera} V.,   {Bulik} T.,  2002, \mn@doi [\apj]
  {10.1086/340304}, \href
  {https://ui.adsabs.harvard.edu/abs/2002ApJ...572..407B} {572, 407}

\bibitem[\protect\citeauthoryear{{Bianchini}, {van de Ven}, {Norris},
  {Schinnerer}  \& {Varri}}{{Bianchini} et~al.}{2016a}]{bianchini_2016b}
{Bianchini} P.,  {van de Ven} G.,  {Norris} M.~A.,  {Schinnerer} E.,   {Varri}
  A.~L.,  2016a, \mn@doi [\mnras] {10.1093/mnras/stw552}, \href
  {https://ui.adsabs.harvard.edu/abs/2016MNRAS.458.3644B} {458, 3644}

\bibitem[\protect\citeauthoryear{{Bianchini}, {Norris}, {van de Ven},
  {Schinnerer}, {Bellini}, {van der Marel}, {Watkins}  \&
  {Anderson}}{{Bianchini} et~al.}{2016b}]{bianchini_2016}
{Bianchini} P.,  {Norris} M.~A.,  {van de Ven} G.,  {Schinnerer} E.,  {Bellini}
  A.,  {van der Marel} R.~P.,  {Watkins} L.~L.,   {Anderson} J.,  2016b,
  \mn@doi [\apjl] {10.3847/2041-8205/820/1/L22}, \href
  {https://ui.adsabs.harvard.edu/abs/2016ApJ...820L..22B} {820, L22}

\bibitem[\protect\citeauthoryear{{Breen} \& {Heggie}}{{Breen} \&
  {Heggie}}{2013}]{breen_2013}
{Breen} P.~G.,  {Heggie} D.~C.,  2013, \mn@doi [\mnras]
  {10.1093/mnras/stt1599}, \href
  {https://ui.adsabs.harvard.edu/abs/2013MNRAS.436..584B} {436, 584}

\bibitem[\protect\citeauthoryear{{Chatterjee}, {Fregeau}, {Umbreit}  \&
  {Rasio}}{{Chatterjee} et~al.}{2010}]{chatterjee_2010}
{Chatterjee} S.,  {Fregeau} J.~M.,  {Umbreit} S.,   {Rasio} F.~A.,  2010,
  \mn@doi [\apj] {10.1088/0004-637X/719/1/915}, \href
  {https://ui.adsabs.harvard.edu/abs/2010ApJ...719..915C} {719, 915}

\bibitem[\protect\citeauthoryear{{Chatterjee}, {Rasio}, {Sills}  \&
  {Glebbeek}}{{Chatterjee} et~al.}{2013}]{chatterjee_2013}
{Chatterjee} S.,  {Rasio} F.~A.,  {Sills} A.,   {Glebbeek} E.,  2013, \mn@doi
  [\apj] {10.1088/0004-637X/777/2/106}, \href
  {https://ui.adsabs.harvard.edu/abs/2013ApJ...777..106C} {777, 106}

\bibitem[\protect\citeauthoryear{{Cheng}, {Li}, {Li}, {Xu}  \& {Fang}}{{Cheng}
  et~al.}{2019a}]{Cheng_2019a}
{Cheng} Z.,  {Li} Z.,  {Li} X.,  {Xu} X.,   {Fang} T.,  2019a, \mn@doi [\apj]
  {10.3847/1538-4357/ab1593}, \href
  {https://ui.adsabs.harvard.edu/abs/2019ApJ...876...59C} {876, 59}

\bibitem[\protect\citeauthoryear{{Cheng}, {Li}, {Fang}, {Li}  \& {Xu}}{{Cheng}
  et~al.}{2019b}]{cheng_2019b}
{Cheng} Z.,  {Li} Z.,  {Fang} T.,  {Li} X.,   {Xu} X.,  2019b, \mn@doi [\apj]
  {10.3847/1538-4357/ab3c6d}, \href
  {https://ui.adsabs.harvard.edu/abs/2019ApJ...883...90C} {883, 90}

\bibitem[\protect\citeauthoryear{{Ferraro} et~al.,}{{Ferraro}
  et~al.}{2012}]{ferraro_2012}
{Ferraro} F.~R.,  et~al., 2012, \mn@doi [\nat] {10.1038/nature11686}, \href
  {https://ui.adsabs.harvard.edu/abs/2012Natur.492..393F} {492, 393}

\bibitem[\protect\citeauthoryear{{Ferraro} et~al.,}{{Ferraro}
  et~al.}{2018}]{ferraro_2018}
{Ferraro} F.~R.,  et~al., 2018, \mn@doi [\apj] {10.3847/1538-4357/aac01c},
  \href {https://ui.adsabs.harvard.edu/abs/2018ApJ...860...36F} {860, 36}

\bibitem[\protect\citeauthoryear{{Fragione} \& {Gualandris}}{{Fragione} \&
  {Gualandris}}{2019}]{fragione_2019}
{Fragione} G.,  {Gualandris} A.,  2019, \mn@doi [\mnras]
  {10.1093/mnras/stz2451}, \href
  {https://ui.adsabs.harvard.edu/abs/2019MNRAS.489.4543F} {489, 4543}

\bibitem[\protect\citeauthoryear{{Fregeau}, {Cheung}, {Portegies Zwart}  \&
  {Rasio}}{{Fregeau} et~al.}{2004}]{fregeau_2004}
{Fregeau} J.~M.,  {Cheung} P.,  {Portegies Zwart} S.~F.,   {Rasio} F.~A.,
  2004, \mn@doi [\mnras] {10.1111/j.1365-2966.2004.07914.x}, \href
  {https://ui.adsabs.harvard.edu/abs/2004MNRAS.352....1F} {352, 1}

\bibitem[\protect\citeauthoryear{{Giersz}, {Heggie}, {Hurley}  \&
  {Hypki}}{{Giersz} et~al.}{2013}]{giersz_2013}
{Giersz} M.,  {Heggie} D.~C.,  {Hurley} J.~R.,   {Hypki} A.,  2013, \mn@doi
  [\mnras] {10.1093/mnras/stt307}, \href
  {https://ui.adsabs.harvard.edu/abs/2013MNRAS.431.2184G} {431, 2184}

\bibitem[\protect\citeauthoryear{{Giersz}, {Leigh}, {Hypki}, {L{\"u}tzgendorf}
  \& {Askar}}{{Giersz} et~al.}{2015}]{giersz_2015}
{Giersz} M.,  {Leigh} N.,  {Hypki} A.,  {L{\"u}tzgendorf} N.,   {Askar} A.,
  2015, \mn@doi [\mnras] {10.1093/mnras/stv2162}, \href
  {https://ui.adsabs.harvard.edu/abs/2015MNRAS.454.3150G} {454, 3150}

\bibitem[\protect\citeauthoryear{{Giesers} et~al.,}{{Giesers}
  et~al.}{2019}]{giesers_2019}
{Giesers} B.,  et~al., 2019, \mn@doi [\aap] {10.1051/0004-6361/201936203},
  \href {https://ui.adsabs.harvard.edu/abs/2019A&A...632A...3G} {632, A3}

\bibitem[\protect\citeauthoryear{{Gill}, {Trenti}, {Miller}, {van der Marel},
  {Hamilton}  \& {Stiavelli}}{{Gill} et~al.}{2008}]{gill_2018}
{Gill} M.,  {Trenti} M.,  {Miller} M.~C.,  {van der Marel} R.,  {Hamilton} D.,
   {Stiavelli} M.,  2008, \mn@doi [\apj] {10.1086/591269}, \href
  {https://ui.adsabs.harvard.edu/abs/2008ApJ...686..303G} {686, 303}

\bibitem[\protect\citeauthoryear{{Gonz{\'a}lez}, {Kremer}, {Chatterjee},
  {Fragione}, {Rodriguez}, {Weatherford}, {Ye}  \& {Rasio}}{{Gonz{\'a}lez}
  et~al.}{2021}]{gonzalez_2021}
{Gonz{\'a}lez} E.,  {Kremer} K.,  {Chatterjee} S.,  {Fragione} G.,  {Rodriguez}
  C.~L.,  {Weatherford} N.~C.,  {Ye} C.~S.,   {Rasio} F.~A.,  2021, \mn@doi
  [\apjl] {10.3847/2041-8213/abdf5b}, \href
  {https://ui.adsabs.harvard.edu/abs/2021ApJ...908L..29G} {908, L29}

\bibitem[\protect\citeauthoryear{Haiman}{Haiman}{2013}]{haiman_2013}
Haiman Z.,  2013, The Formation of the First Massive Black Holes.
Springer, Berlin, Heidelberg, pp 293--341, \mn@doi{10.1007/978-3-642-32362-1_6}

\bibitem[\protect\citeauthoryear{{Harris}}{{Harris}}{1996}]{harris_1996}
{Harris} W.~E.,  1996, \mn@doi [\aj] {10.1086/118116}, \href
  {https://ui.adsabs.harvard.edu/abs/1996AJ....112.1487H} {112, 1487}

\bibitem[\protect\citeauthoryear{{Heggie}, {Trenti}  \& {Hut}}{{Heggie}
  et~al.}{2006}]{heggie_2006}
{Heggie} D.~C.,  {Trenti} M.,   {Hut} P.,  2006, \mn@doi [\mnras]
  {10.1111/j.1365-2966.2006.10122.x}, \href
  {https://ui.adsabs.harvard.edu/abs/2006MNRAS.368..677H} {368, 677}

\bibitem[\protect\citeauthoryear{{H{\'e}non}}{{H{\'e}non}}{1971a}]{henon_1971b}
{H{\'e}non} M.,  1971a, \mn@doi [\apss] {10.1007/BF00649159}, \href
  {https://ui.adsabs.harvard.edu/abs/1971Ap&SS..13..284H} {13, 284}

\bibitem[\protect\citeauthoryear{{H{\'e}non}}{{H{\'e}non}}{1971b}]{henon_1971a}
{H{\'e}non} M.~H.,  1971b, \mn@doi [\apss] {10.1007/BF00649201}, \href
  {https://ui.adsabs.harvard.edu/abs/1971Ap&SS..14..151H} {14, 151}

\bibitem[\protect\citeauthoryear{{Hills}}{{Hills}}{1988}]{hills_1988}
{Hills} J.~G.,  1988, \mn@doi [\nat] {10.1038/331687a0}, \href
  {https://ui.adsabs.harvard.edu/abs/1988Natur.331..687H} {331, 687}

\bibitem[\protect\citeauthoryear{{Hobbs}, {Lorimer}, {Lyne}  \&
  {Kramer}}{{Hobbs} et~al.}{2005}]{hobbs_2005}
{Hobbs} G.,  {Lorimer} D.~R.,  {Lyne} A.~G.,   {Kramer} M.,  2005, \mn@doi
  [\mnras] {10.1111/j.1365-2966.2005.09087.x}, \href
  {https://ui.adsabs.harvard.edu/abs/2005MNRAS.360..974H} {360, 974}

\bibitem[\protect\citeauthoryear{{Hong}, {Vesperini}, {Sollima}, {McMillan},
  {D'Antona}  \& {D'Ercole}}{{Hong} et~al.}{2015}]{hong_2015}
{Hong} J.,  {Vesperini} E.,  {Sollima} A.,  {McMillan} S. L.~W.,  {D'Antona}
  F.,   {D'Ercole} A.,  2015, \mn@doi [\mnras] {10.1093/mnras/stv306}, \href
  {https://ui.adsabs.harvard.edu/abs/2015MNRAS.449..629H} {449, 629}

\bibitem[\protect\citeauthoryear{{Hong}, {Askar}, {Giersz}, {Hypki}  \&
  {Yoon}}{{Hong} et~al.}{2020}]{hong_2020}
{Hong} J.,  {Askar} A.,  {Giersz} M.,  {Hypki} A.,   {Yoon} S.-J.,  2020,
  \mn@doi [\mnras] {10.1093/mnras/staa2677}, \href
  {https://ui.adsabs.harvard.edu/abs/2020MNRAS.498.4287H} {498, 4287}

\bibitem[\protect\citeauthoryear{{Hunter}}{{Hunter}}{2007}]{hunter_2007}
{Hunter} J.~D.,  2007, \mn@doi [Computing in Science and Engineering]
  {10.1109/MCSE.2007.55}, \href
  {https://ui.adsabs.harvard.edu/abs/2007CSE.....9...90H} {9, 90}

\bibitem[\protect\citeauthoryear{{Hurley}, {Tout}  \& {Pols}}{{Hurley}
  et~al.}{2002}]{hurley_2002}
{Hurley} J.~R.,  {Tout} C.~A.,   {Pols} O.~R.,  2002, \mn@doi [\mnras]
  {10.1046/j.1365-8711.2002.05038.x}, \href
  {https://ui.adsabs.harvard.edu/abs/2002MNRAS.329..897H} {329, 897}

\bibitem[\protect\citeauthoryear{{Hurley}, {Aarseth}  \& {Shara}}{{Hurley}
  et~al.}{2007}]{hurley_2007}
{Hurley} J.~R.,  {Aarseth} S.~J.,   {Shara} M.~M.,  2007, \mn@doi [\apj]
  {10.1086/517879}, \href
  {https://ui.adsabs.harvard.edu/abs/2007ApJ...665..707H} {665, 707}

\bibitem[\protect\citeauthoryear{{Hypki} \& {Giersz}}{{Hypki} \&
  {Giersz}}{2013}]{hypki_2013}
{Hypki} A.,  {Giersz} M.,  2013, \mn@doi [\mnras] {10.1093/mnras/sts415}, \href
  {https://ui.adsabs.harvard.edu/abs/2013MNRAS.429.1221H} {429, 1221}

\bibitem[\protect\citeauthoryear{{Ji} \& {Bregman}}{{Ji} \&
  {Bregman}}{2015}]{ji_2015}
{Ji} J.,  {Bregman} J.~N.,  2015, \mn@doi [\apj] {10.1088/0004-637X/807/1/32},
  \href {https://ui.adsabs.harvard.edu/abs/2015ApJ...807...32J} {807, 32}

\bibitem[\protect\citeauthoryear{{Kaderali}, {Hunt}, {Webb}, {Price-Jones}  \&
  {Carlberg}}{{Kaderali} et~al.}{2019}]{kaderali_2019}
{Kaderali} S.,  {Hunt} J. A.~S.,  {Webb} J.~J.,  {Price-Jones} N.,   {Carlberg}
  R.,  2019, \mn@doi [\mnras] {10.1093/mnrasl/slz015}, \href
  {https://ui.adsabs.harvard.edu/abs/2019MNRAS.484L.114K} {484, L114}

\bibitem[\protect\citeauthoryear{{Kamann}, {Wisotzki}, {Roth}, {Gerssen},
  {Husser}, {Sandin}  \& {Weilbacher}}{{Kamann} et~al.}{2014}]{kamann_2014}
{Kamann} S.,  {Wisotzki} L.,  {Roth} M.~M.,  {Gerssen} J.,  {Husser} T.~O.,
  {Sandin} C.,   {Weilbacher} P.,  2014, \mn@doi [\aap]
  {10.1051/0004-6361/201322183}, \href
  {https://ui.adsabs.harvard.edu/abs/2014A&A...566A..58K} {566, A58}

\bibitem[\protect\citeauthoryear{{Kamann} et~al.,}{{Kamann}
  et~al.}{2016}]{kamann_2016}
{Kamann} S.,  et~al., 2016, \mn@doi [\aap] {10.1051/0004-6361/201527065}, \href
  {https://ui.adsabs.harvard.edu/abs/2016A&A...588A.149K} {588, A149}

\bibitem[\protect\citeauthoryear{{Kamann} et~al.,}{{Kamann}
  et~al.}{2018}]{kamann_2018}
{Kamann} S.,  et~al., 2018, \mn@doi [\mnras] {10.1093/mnras/stx2719}, \href
  {https://ui.adsabs.harvard.edu/abs/2018MNRAS.473.5591K} {473, 5591}

\bibitem[\protect\citeauthoryear{{King}}{{King}}{1962}]{king_1962}
{King} I.,  1962, \mn@doi [\aj] {10.1086/108756}, \href
  {https://ui.adsabs.harvard.edu/abs/1962AJ.....67..471K} {67, 471}

\bibitem[\protect\citeauthoryear{{Kroupa}}{{Kroupa}}{2001}]{kroupa_2001}
{Kroupa} P.,  2001, \mn@doi [\mnras] {10.1046/j.1365-8711.2001.04022.x}, \href
  {https://ui.adsabs.harvard.edu/abs/2001MNRAS.322..231K} {322, 231}

\bibitem[\protect\citeauthoryear{{Kunder} et~al.,}{{Kunder}
  et~al.}{2017}]{kunder_2017}
{Kunder} A.,  et~al., 2017, \mn@doi [\aj] {10.3847/1538-3881/153/2/75}, \href
  {https://ui.adsabs.harvard.edu/abs/2017AJ....153...75K} {153, 75}

\bibitem[\protect\citeauthoryear{{Lanzoni} et~al.,}{{Lanzoni}
  et~al.}{2013}]{lanzoni_2013}
{Lanzoni} B.,  et~al., 2013, \mn@doi [\apj] {10.1088/0004-637X/769/2/107},
  \href {https://ui.adsabs.harvard.edu/abs/2013ApJ...769..107L} {769, 107}

\bibitem[\protect\citeauthoryear{{Libralato} et~al.,}{{Libralato}
  et~al.}{2018}]{libralato_2018}
{Libralato} M.,  et~al., 2018, \mn@doi [\apj] {10.3847/1538-4357/aac6c0}, \href
  {https://ui.adsabs.harvard.edu/abs/2018ApJ...861...99L} {861, 99}

\bibitem[\protect\citeauthoryear{{L{\"u}tzgendorf} et~al.,}{{L{\"u}tzgendorf}
  et~al.}{2013}]{lutzgendorf_2013}
{L{\"u}tzgendorf} N.,  et~al., 2013, \mn@doi [\aap]
  {10.1051/0004-6361/201220307}, \href
  {http://adsabs.harvard.edu/abs/2013A%26A...552A..49L} {552, A49}

\bibitem[\protect\citeauthoryear{{Mackey}, {Wilkinson}, {Davies}  \&
  {Gilmore}}{{Mackey} et~al.}{2008}]{mackey_2008}
{Mackey} A.~D.,  {Wilkinson} M.~I.,  {Davies} M.~B.,   {Gilmore} G.~F.,  2008,
  \mn@doi [\mnras] {10.1111/j.1365-2966.2008.13052.x}, \href
  {https://ui.adsabs.harvard.edu/abs/2008MNRAS.386...65M} {386, 65}

\bibitem[\protect\citeauthoryear{{Mann} et~al.,}{{Mann}
  et~al.}{2019}]{mann_2019}
{Mann} C.~R.,  et~al., 2019, \mn@doi [\apj] {10.3847/1538-4357/ab0e6d}, \href
  {https://ui.adsabs.harvard.edu/abs/2019ApJ...875....1M} {875, 1}

\bibitem[\protect\citeauthoryear{{Mapelli}, {Sigurdsson}, {Ferraro}, {Colpi},
  {Possenti}  \& {Lanzoni}}{{Mapelli} et~al.}{2006}]{mapelli_2006}
{Mapelli} M.,  {Sigurdsson} S.,  {Ferraro} F.~R.,  {Colpi} M.,  {Possenti} A.,
   {Lanzoni} B.,  2006, \mn@doi [\mnras] {10.1111/j.1365-2966.2006.11038.x},
  \href {https://ui.adsabs.harvard.edu/abs/2006MNRAS.373..361M} {373, 361}

\bibitem[\protect\citeauthoryear{{McNamara}, {Harrison}, {Baumgardt}  \&
  {Khalaj}}{{McNamara} et~al.}{2012}]{mcnamara_2012}
{McNamara} B.~J.,  {Harrison} T.~E.,  {Baumgardt} H.,   {Khalaj} P.,  2012,
  \mn@doi [\apj] {10.1088/0004-637X/745/2/175}, \href
  {https://ui.adsabs.harvard.edu/abs/2012ApJ...745..175M} {745, 175}

\bibitem[\protect\citeauthoryear{{Milone} et~al.,}{{Milone}
  et~al.}{2012}]{milone_2012}
{Milone} A.~P.,  et~al., 2012, \mn@doi [\aap] {10.1051/0004-6361/201016384},
  \href {https://ui.adsabs.harvard.edu/abs/2012A&A...540A..16M} {540, A16}

\bibitem[\protect\citeauthoryear{{Milone} et~al.,}{{Milone}
  et~al.}{2020}]{milone_2020}
{Milone} A.~P.,  et~al., 2020, \mn@doi [\mnras] {10.1093/mnras/stz3629}, \href
  {https://ui.adsabs.harvard.edu/abs/2020MNRAS.492.5457M} {492, 5457}

\bibitem[\protect\citeauthoryear{{Morscher}, {Pattabiraman}, {Rodriguez},
  {Rasio}  \& {Umbreit}}{{Morscher} et~al.}{2015}]{morscher_2015}
{Morscher} M.,  {Pattabiraman} B.,  {Rodriguez} C.,  {Rasio} F.~A.,   {Umbreit}
  S.,  2015, \mn@doi [\apj] {10.1088/0004-637X/800/1/9}, \href
  {https://ui.adsabs.harvard.edu/abs/2015ApJ...800....9M} {800, 9}

\bibitem[\protect\citeauthoryear{{Noyola}, {Gebhardt}  \& {Bergmann}}{{Noyola}
  et~al.}{2008}]{noyola_2008}
{Noyola} E.,  {Gebhardt} K.,   {Bergmann} M.,  2008, \mn@doi [\apj]
  {10.1086/529002}, \href {http://adsabs.harvard.edu/abs/2008ApJ...676.1008N}
  {676, 1008}

\bibitem[\protect\citeauthoryear{{Portegies Zwart}, {Baumgardt}, {Hut},
  {Makino}  \& {McMillan}}{{Portegies Zwart}
  et~al.}{2004}]{portegies_zwart_2004}
{Portegies Zwart} S.~F.,  {Baumgardt} H.,  {Hut} P.,  {Makino} J.,   {McMillan}
  S. L.~W.,  2004, \mn@doi [\nat] {10.1038/nature02448}, \href
  {https://ui.adsabs.harvard.edu/abs/2004Natur.428..724P} {428, 724}

\bibitem[\protect\citeauthoryear{{Rizzuto} et~al.,}{{Rizzuto}
  et~al.}{2021}]{rizzuto_2021}
{Rizzuto} F.~P.,  et~al., 2021, \mn@doi [\mnras] {10.1093/mnras/staa3634},
  \href {https://ui.adsabs.harvard.edu/abs/2021MNRAS.501.5257R} {501, 5257}

\bibitem[\protect\citeauthoryear{{Sippel} \& {Hurley}}{{Sippel} \&
  {Hurley}}{2013}]{sipel_2013}
{Sippel} A.~C.,  {Hurley} J.~R.,  2013, \mn@doi [\mnras]
  {10.1093/mnrasl/sls044}, \href
  {https://ui.adsabs.harvard.edu/abs/2013MNRAS.430L..30S} {430, L30}

\bibitem[\protect\citeauthoryear{{Sollima}}{{Sollima}}{2020}]{sollima_2020}
{Sollima} A.,  2020, \mn@doi [\mnras] {10.1093/mnras/staa1209}, \href
  {https://ui.adsabs.harvard.edu/abs/2020MNRAS.495.2222S} {495, 2222}

\bibitem[\protect\citeauthoryear{{Sollima}, {Beccari}, {Ferraro}, {Fusi Pecci}
  \& {Sarajedini}}{{Sollima} et~al.}{2007}]{sollima_2007}
{Sollima} A.,  {Beccari} G.,  {Ferraro} F.~R.,  {Fusi Pecci} F.,   {Sarajedini}
  A.,  2007, \mn@doi [\mnras] {10.1111/j.1365-2966.2007.12116.x}, \href
  {https://ui.adsabs.harvard.edu/abs/2007MNRAS.380..781S} {380, 781}

\bibitem[\protect\citeauthoryear{{Spitzer}}{{Spitzer}}{1969}]{spitzer_1969}
{Spitzer} Lyman J.,  1969, \mn@doi [\apjl] {10.1086/180451}, \href
  {https://ui.adsabs.harvard.edu/abs/1969ApJ...158L.139S} {158, L139}

\bibitem[\protect\citeauthoryear{{Spitzer}}{{Spitzer}}{1987}]{spitzer_1987}
{Spitzer} L.,  1987, {Dynamical evolution of globular clusters}.
Princeton University Press

\bibitem[\protect\citeauthoryear{{Trenti} \& {van der Marel}}{{Trenti} \& {van
  der Marel}}{2013}]{trenti_2013}
{Trenti} M.,  {van der Marel} R.,  2013, \mn@doi [\mnras]
  {10.1093/mnras/stt1521}, \href
  {https://ui.adsabs.harvard.edu/abs/2013MNRAS.435.3272T} {435, 3272}

\bibitem[\protect\citeauthoryear{{Trenti}, {Ardi}, {Mineshige}  \&
  {Hut}}{{Trenti} et~al.}{2007}]{trenti_2007}
{Trenti} M.,  {Ardi} E.,  {Mineshige} S.,   {Hut} P.,  2007, \mn@doi [\mnras]
  {10.1111/j.1365-2966.2006.11189.x}, \href
  {https://ui.adsabs.harvard.edu/abs/2007MNRAS.374..857T} {374, 857}

\bibitem[\protect\citeauthoryear{{Vitral} \& {Mamon}}{{Vitral} \&
  {Mamon}}{2021}]{vitral_2021}
{Vitral} E.,  {Mamon} G.~A.,  2021, \mn@doi [\aap]
  {10.1051/0004-6361/202039650}, \href
  {https://ui.adsabs.harvard.edu/abs/2021A&A...646A..63V} {646, A63}

\bibitem[\protect\citeauthoryear{{Wang} et~al.,}{{Wang}
  et~al.}{2016}]{wang_2016}
{Wang} L.,  et~al., 2016, \mn@doi [\mnras] {10.1093/mnras/stw274}, \href
  {https://ui.adsabs.harvard.edu/abs/2016MNRAS.458.1450W} {458, 1450}

\bibitem[\protect\citeauthoryear{{Weatherford}, {Chatterjee}, {Rodriguez}  \&
  {Rasio}}{{Weatherford} et~al.}{2018}]{weatherford_2018}
{Weatherford} N.~C.,  {Chatterjee} S.,  {Rodriguez} C.~L.,   {Rasio} F.~A.,
  2018, \mn@doi [\apj] {10.3847/1538-4357/aad63d}, \href
  {https://ui.adsabs.harvard.edu/abs/2018ApJ...864...13W} {864, 13}

\bibitem[\protect\citeauthoryear{{Weatherford}, {Chatterjee}, {Kremer}  \&
  {Rasio}}{{Weatherford} et~al.}{2020}]{weatherford_2020}
{Weatherford} N.~C.,  {Chatterjee} S.,  {Kremer} K.,   {Rasio} F.~A.,  2020,
  \mn@doi [\apj] {10.3847/1538-4357/ab9f98}, \href
  {https://ui.adsabs.harvard.edu/abs/2020ApJ...898..162W} {898, 162}

\bibitem[\protect\citeauthoryear{{Zocchi}, {Gieles}  \&
  {H{\'e}nault-Brunet}}{{Zocchi} et~al.}{2019}]{zocchi_2019}
{Zocchi} A.,  {Gieles} M.,   {H{\'e}nault-Brunet} V.,  2019, \mn@doi [\mnras]
  {10.1093/mnras/sty1508}, \href
  {http://adsabs.harvard.edu/abs/2019MNRAS.482.4713Z} {482, 4713}

\bibitem[\protect\citeauthoryear{{de Vita}, {Trenti}, {Bianchini}, {Askar},
  {Giersz}  \& {van de Ven}}{{de Vita} et~al.}{2017}]{de_vita_2017}
{de Vita} R.,  {Trenti} M.,  {Bianchini} P.,  {Askar} A.,  {Giersz} M.,   {van
  de Ven} G.,  2017, \mn@doi [\mnras] {10.1093/mnras/stx325}, \href
  {https://ui.adsabs.harvard.edu/abs/2017MNRAS.467.4057D} {467, 4057}

\bibitem[\protect\citeauthoryear{{{\v{S}}ubr}, {Fragione}  \&
  {Dabringhausen}}{{{\v{S}}ubr} et~al.}{2019}]{subr_2019}
{{\v{S}}ubr} L.,  {Fragione} G.,   {Dabringhausen} J.,  2019, \mn@doi [\mnras]
  {10.1093/mnras/stz162}, \href
  {https://ui.adsabs.harvard.edu/abs/2019MNRAS.484.2974S} {484, 2974}

\bibitem[\protect\citeauthoryear{{van der Marel} \& {Anderson}}{{van der Marel}
  \& {Anderson}}{2010}]{van_der_marel_2010}
{van der Marel} R.~P.,  {Anderson} J.,  2010, \mn@doi [\apj]
  {10.1088/0004-637X/710/2/1063}, \href
  {http://adsabs.harvard.edu/abs/2010ApJ...710.1063V} {710, 1063}

\bibitem[\protect\citeauthoryear{{van der Walt}, {Colbert}  \&
  {Varoquaux}}{{van der Walt} et~al.}{2011}]{van_der_walt_2011}
{van der Walt} S.,  {Colbert} S.~C.,   {Varoquaux} G.,  2011, \mn@doi
  [Computing in Science and Engineering] {10.1109/MCSE.2011.37}, \href
  {https://ui.adsabs.harvard.edu/abs/2011CSE....13b..22V} {13, 22}

\makeatother
\end{thebibliography}

%\begin{thebibliography}{99}
%\end{thebibliography}

%%%%%%%%%%%%%%%%%%%%%%%%%%%%%%%%%%%%%%%%%%%%%%%%%%

%%%%%%%%%%%%%%%%% APPENDICES %%%%%%%%%%%%%%%%%%%%%

\appendix

\section{Initial Conditions}
\label{secA:initial_condition}
All GCs in the \textsc{MOCCA}-Survey Database I have an initial mass function given by \cite{kroupa_2001}, with stellar masses ranging from  $0.08\,M_{\odot}$ to $100\,M_{\odot}$. Whereas the full survey has a more extensive parameter space for initial conditions, our sample of GCs with initial binary fraction $f_{\text{bin}}=10\%$ has a combination of initial conditions given by Table \ref{tab:init_cond}. For further details on the \textsc{MOCCA}-Survey Database I, we refer the reader to \cite{askar_2017a}.

In the simulated GCs, the velocity of stellar-mass black holes due to natal kicks can follow two distinct approaches. If the cluster has a “no Fallback” approach, then the natal kick velocities come from a Maxwellian velocity distribution with $\sigma=265\,\text{km/s}$ \citep{hobbs_2005}. Alternatively, clusters can follow a “Fallback” approach, where stellar-mass black holes can have their masses and natal kick velocities modified by the mass fallback prescription provided by \cite{belczynski_2002}. The latter allows for a stellar-mass black hole retention fraction of $15\%$ to $55\%$ within the first $30\,\text{Myr}$ \citep{askar_2018b}.

\begin{table}
    \centering
    \caption{Initial conditions for the simulated GCs in our sample. $N$ is the initial number of objects (single+binaries), $r_{\text{t}}$ is the tidal radius, $W_0$ is the King parameter and $Z$ the metallicity. Only one cluster is tidally filling. Natal kicks for neutron stars and stellar-mass black holes follow a mass fallback model \protect\citep{belczynski_2002} or a Maxwellian distribution \protect\citep{hobbs_2005}.}
    \begin{tabular}{l|l}
    \hline
    Parameter & values \\
    \hline
    $N$    & $400000$, $700000$, $1200000$  \\
    $r_{\text{t}}$ [pc]   & 30.0, 60.0, 120.0  \\
    $r_{\text{t}}/r_{\text{h}}$ & 25.0, 50., filling \\
    $W_0$ & 3.0, 6.0, 9.0 \\
    $Z$ & 0.001, 0.005, 0.02 \\
    Natal kicks & Fallback and no Fallback\\
    \hline
    \end{tabular}
    \label{tab:init_cond}
\end{table}

\section{Observational errors}
\label{secA:errors}
In this study we have included observational errors and noise to the kinematics. For each simulated cluster we have the 3-dimensional velocities to which we add a noise accordingly the observational error expected for either the line-of-sight velocity or the proper motions. In both cases the `observed' velocity will be given by:
\begin{equation}
    \text{v}_{\text{obs}} = \text{v}_{\text{sim}} + N(0,\delta^2)\,,
\end{equation}
where $\text{v}_{\text{sim}}$ is the velocity from the simulation and $N(0,\delta^2)$ is a value randomly sampled form a Gaussian distribution centred in $0$ with dispersion given by the observational error $\delta$. The latter serve as a noise due to the observational errors.

\begin{figure}
    \centering
    \includegraphics[width=1\linewidth]{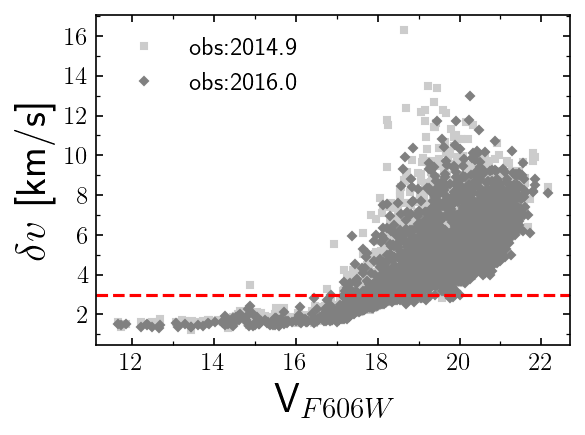}
    \caption{Observational errors of line-of-sight radial velocities for NGC3201 from \protect\cite{giesers_2019}. We selected the two epochs with the largest sample of observations. The red dashed line represents the error limit for kinematics in this work, assuming a maximum error of $\delta_{v}\sim3\,\text{km/s}$.}
    \label{fig:vz_error}
\end{figure}

In the case of the line-of-sight velocities we use the observational errors of MUSE/VLT data from \cite{giesers_2019}. We bin the observed stars by magnitude and get the median error in each magnitude bin, Figure \ref{fig:vz_error} shows the distribution of errors and their median value. We use the V magnitude of each star in the simulation to assign an error and scatter, assuming the cluster is at a distance of $5\,\text{kpc}$.  

We use a fixed error of $\delta_{\text{pm}}=0.1\,\text{mas/yr}$, taken from \cite{libralato_2018} for improved HST astrometry. We transform this value to $\text{km/s}$ assuming the clusters are at a distance of $5\,\text{kpc}$. 

As we added a noise to the observed kinematics we use the following likelihood approach to obtain the intrinsic velocity dispersion $\sigma$ and mean velocitiy $\overline{\text{v}}$:

\begin{equation}
\mathcal{L}(\sigma,\overline{\text{v}}|\vec{\text{v}}_{\text{obs}})=\prod_{i=0}^{N}
\frac{1}{\sqrt{2\pi(\sigma^2+\delta_i^2)}}\exp\left(\frac{(\text{v}_{\text{obs},i}-\overline{\text{v}})^2}{2(\sigma^2+\delta_i^2)}\right)
\end{equation}

%\section{Probability of a being a binary}
%Additional information on the probability of being a binary. Describe some of the properties of the detected binaries. Period, Semi-major axis and inclination. 

%\begin{figure}
%    \centering
%    \includegraphics[width=0.8\linewidth]{FIGURES/probability_distribution.pdf}
%    \caption{Caption}
%    \label{fig:my_label}
%\end{figure}

%\begin{figure}
%    \centering
%    \includegraphics[width=0.9\linewidth]{FIGURES/detected_binaries.pdf}
%    \caption{Detected binary fraction with respect to the real binary fraction. We find two distinctive groups in our sample of simulated GCs, for the ones in the bottom left we find most binaries using variations in RVs. As it will be discussed in Section \protect\ref{sec:imbh_bin}, these GCs have an IMBH in their centre. For the second group in the middle right we systematically underestimate the binary fraction, which could be due the fraction of long period binaries which where not detected.[Check which is the maximum period for binaries in each simulated GC.]}
%    \label{fig:detected_binaries}
%\end{figure}

\section{Gobular clusters hosting a black hole subsystem}
\label{secA:table}
In Figure \ref{fig:bin_frac_milone} we show the binary fraction within the core radius and outside the half-mass radius for a sample of Galactic GCs \citep{milone_2012}. We have cross-matched these clusters with the list of candidate Galactic GCs from \cite{askar_2018b} and \cite{weatherford_2020} to find those which have retained stellar mass black holes (BHs). We notice that indeed the candidate GCs fall closer to the 1-to-1 ratio of binary fractions (see Figure \ref{fig:bin_frac_milone}). We define an orthogonal distance to the 1-to-1 line, for each cluster in the sample, as: 
\begin{equation}
    \Delta = \frac{f_{\text{bin}}(R<R_{\text{c}}) - f_{\text{bin}}(R>R_{\text{hm}})}{\sqrt{2}}\,.
\end{equation}

In Table \ref{tab:bin_frac} we summarize the clusters used in Figure \ref{fig:bin_frac_milone}, indicating their names, binary fractions (core $f_{\text{c}}$ and half-mass $f_{\text{hm}}$) from \cite{milone_2012}, the distance from the 1-to-1 relation $\Delta$. We have highlighted in bold the clusters that have an off-set of $\Delta<5$, which is below the median value for the distances to the 1-to-1 relation. We also include the number of retained BHs ($N_{\text{BHs}}$) and the total mass of the retained BHs ($M_{\text{BHS}}$) from both \cite{askar_2018b} and \cite{weatherford_2020}. We include estimated mass of a central IMBH for three GCs in the sample with possible IMBHs in their centre.

\begin{table*}
    \centering
    \caption{Galactic globular clusters in Figure \protect\ref{fig:bin_frac_milone}. For each cluster we show its name (GC), the binary fractions within the core radius $f_{\text{c}}$ and outside the half-mass radius $f_{\text{hm}}$ from \protect\cite{milone_2012}$^{(a)}$. We calculated the distance $\Delta$ between each  GC and the 1-to-1 line, and highlight those which have $\Delta<5$. For the clusters from \protect\cite{askar_2018b}$^{(b)}$ and \protect\cite{weatherford_2020}$^{(c)}$ we include the estimated number of retained BHs $N_{\text{BHs}}$ and the total mass in BHs $M_{\text{BHs}}$. IMBH masses ($M_{\text{IMBH}}$) come from \protect\cite{kamann_2014}$^{(d)}$ and \protect\cite{kamann_2016}$^{(e)}$. } 
    \begin{tabular}{|l|r|r|r|r|r|r|r|r|}
    \hline
    GC  & $f_{\text{c}}^{(a)}\,[\%]$ & $f_{\text{hm}}^{(a)}\,[\%]$ & $\Delta$ & $N_{\text{BHs}}^{(b)}$ & $M_{\text{BHs}}^{(b)} [M_{\odot}]$ & $N_{\text{BHs}}^{(c)}$ & $M_{\text{BHs}}^{(c)} [M_{\odot}]$ & $M_{\text{IMBH}}\,[M_{\odot}]$\\
    \hline
\textbf{ARP 2} & $ 18.6 \pm  2.0 $ & $ 18.2 \pm  6.2 $ & \textbf{0.3}&    &  &  & \vspace{4pt}\\
       E 3 & $ 72.0 \pm  8.6 $ & $ 16.4 \pm 21.4 $ &       39.3&    &  &  & \vspace{4pt}\\
   NGC 288 & $ 11.2 \pm  1.0 $ & $ 18.4 \pm  8.0 $ &        5.1& $118_{-35}^{+58}$ & $1473.0_{-354}^{+566}$ & $   26^{ -14}_{ +52}$ & $   594^{ -305}_{+1216}$ \vspace{4pt}\\
\textbf{NGC 1261} & $  4.6 \pm  1.8 $ & $  4.0 \pm  0.6 $ & \textbf{0.4}&   &  & $   39^{ -20}_{ +81}$ & $   845^{ -416}_{+1742}$ \vspace{4pt}\\
\textbf{NGC 2298} & $ 15.4 \pm  1.8 $ & $  9.4 \pm  0.8 $ & \textbf{4.2}&   &  & $    1^{  +0}_{  +3}$ & $    21^{   -4}_{  +80}$ \vspace{4pt}\\
  NGC 4147 & $ 26.2 \pm  9.4 $ & $  3.8 \pm  1.2 $ &       15.8&   &  & $    2^{  +0}_{  +9}$ & $    55^{   -9}_{ +235}$ \vspace{4pt}\\
\textbf{NGC 4590} & $ 11.4 \pm  1.2 $ & $ 10.6 \pm  1.4 $ & \textbf{0.6}&   $  71_{-18}^{+29} $ & $ 847.8_{-166}^{260} $ & $   17^{  -6}_{ +38}$ & $   338^{ -103}_{ +888}$ \vspace{4pt}\\
\textbf{NGC 5272} & $  5.4 \pm  1.4 $ & $  3.8 \pm  0.6 $ & \textbf{1.1}&   $  55_{-13}^{+20} $ & $ 632.9_{-109}^{+169} $ & $   25^{  -4}_{+112}$ & $   587^{  -71}_{+2786}$ & $<5.3\times10^3(d)$\vspace{4pt}\\
  NGC 5466 & $ 14.2 \pm  0.8 $ & $  3.2 \pm  7.0 $ &        7.8&   $  191_{-63}^{+110} $ & $ 2512.2_{-703}^{+1165} $ & $   19^{ -10}_{ +67}$ & $   423^{ -197}_{+1740}$ \vspace{4pt}\\
  NGC 5927 & $ 10.4 \pm  1.8 $ & $  1.2 \pm  0.6 $ &        6.5&   &  & $  123^{ -69}_{+273}$ & $  2499^{-1320}_{+6110}$ \vspace{4pt}\\
\textbf{NGC 6101} & $ 10.0 \pm  0.8 $ & $ 10.8 \pm  1.4 $ & \textbf{0.6}&  $  89_{-24}^{+40} $ & $ 1085.6_{-234}^{+370}$ & $  125^{-104}_{+236}$ & $  3051^{-2497}_{+5880}$ \vspace{4pt}\\
  NGC 6144 & $ 13.2 \pm  1.2 $ & $  6.0 \pm  1.4 $ &        5.1&  $  84_{-23}^{+40} $ & $ 1012.2_{-213}^{+335} $ & $   13^{  -7}_{ +36}$ & $   299^{ -144}_{ +855}$ \vspace{4pt}\\
\textbf{NGC 6205} & $  1.0 \pm  0.6 $ & $  2.4 \pm  0.6 $ & \textbf{1.0}&   $  34_{-6}^{+10} $ & $ 366.8_{-46}^{+72} $ & $  128^{ -61}_{+345}$ & $  2786^{-1178}_{+8444}$ & $<8.1\times10^{3}(d)$\vspace{4pt}\\
  NGC 6218 & $ 11.4 \pm  1.0 $ & $  2.2 \pm  2.6 $ &        6.5&    &  & $   22^{ -11}_{ +65}$ & $   509^{ -233}_{+1507}$ \vspace{4pt}\\
\textbf{NGC 6254} & $  7.8 \pm  0.8 $ & $  5.4 \pm  1.4 $ & \textbf{1.7}&     &  & $   30^{ -12}_{ +69}$ & $   622^{ -206}_{+1612}$ \vspace{4pt}\\
  NGC 6352 & $ 18.4 \pm  1.6 $ & $  7.8 \pm  3.4 $ &        7.5&     &  & $   14^{  -5}_{ +39}$ & $   298^{  -98}_{ +875}$ \vspace{4pt}\\
\textbf{NGC 6362} & $ 12.0 \pm  0.8 $ & $  6.4 \pm  7.4 $ & \textbf{4.0}&   $  86_{-23}^{+38} $ & $ 1039.3_{-221}^{+238} $ &  &  \vspace{4pt}\\
\textbf{NGC 6397} & $  7.0 \pm  3.6 $ & $  2.8 \pm  5.2 $ & \textbf{3.0}&     &  & $    3^{  +0}_{ +16}$ & $    72^{   +0}_{ +421}$ & $600\pm200(e)$\vspace{4pt}\\
  NGC 6496 & $ 17.8 \pm  1.2 $ & $  9.2 \pm  4.8 $ &        6.1&    $  58_{-14}^{+22} $ & $ 672.2_{-119}^{+185} $ &  &  \vspace{4pt}\\
\textbf{NGC 6535} & $  9.2 \pm  3.2 $ & $  5.6 \pm  2.0 $ & \textbf{2.5}&    &  & $    1^{  +0}_{  +5}$ & $    24^{   -2}_{ +125}$ \vspace{4pt}\\
\textbf{NGC 6584} & $  9.0 \pm  1.2 $ & $  5.0 \pm  0.6 $ & \textbf{2.8}&   $  40_{-8}^{+13} $ & $ 451.5_{-64}^{+101} $ & $   11^{  -3}_{ +29}$ & $   231^{  -61}_{ +687}$ \vspace{4pt}\\
  NGC 6637 & $ 12.4 \pm  2.0 $ & $  2.6 \pm  0.6 $ &        6.9&     &  & $   58^{ -25}_{+123}$ & $  1154^{ -478}_{+2728}$ \vspace{4pt}\\
  NGC 6652 & $ 34.4 \pm 11.0 $ & $  5.4 \pm  1.2 $ &       20.5&     &  & $    5^{  -1}_{ +22}$ & $   107^{  -13}_{ +501}$ \vspace{4pt}\\
\textbf{NGC 6723} & $  6.2 \pm  0.8 $ & $  3.4 \pm  0.8 $ & \textbf{2.0}&   $  51_{-11}^{+18} $ & $ 577.7_{-95}^{+147} $ & $   60^{ -24}_{+189}$ & $  1243^{ -491}_{+4528}$ \vspace{4pt}\\
\textbf{NGC 6779} & $ 10.0 \pm  1.8 $ & $  4.6 \pm  0.6 $ & \textbf{3.8}&    $  48_{-11}^{+17} $ & $ 543.1_{-141}^{220} $ & $   51^{ -24}_{+103}$ & $  1068^{ -430}_{+2329}$ \vspace{4pt}\\
  NGC 6838 & $ 30.4 \pm  3.4 $ & $ 20.8 \pm  2.8 $ &        6.8&    &  & $   17^{  -6}_{ +60}$ & $   363^{ -119}_{+1446}$ \vspace{4pt}\\
\textbf{NGC 6981} & $  9.8 \pm  1.8 $ & $  6.8 \pm  1.2 $ & \textbf{2.1}&   $  84_{-22}^{+37} $ & $ 1010.6_{-212}^{+334} $ & $   27^{ -17}_{ +61}$ & $   573^{ -334}_{+1747}$ \vspace{4pt}\\
\textbf{NGC 7099} & $  7.0 \pm  3.0 $ & $  2.6 \pm  0.6 $ & \textbf{3.1}&     &  & $    5^{  +0}_{ +28}$ & $   130^{   -1}_{ +714}$ \vspace{4pt}\\
 PALOMAR 1 & $ 66.6 \pm 19.2 $ & $ 19.0 \pm  6.2 $ &       33.7&    &  &  &  \vspace{4pt}\\
PALOMAR 12 & $ 26.0 \pm 11.4 $ & $ 13.2 \pm  3.8 $ &        9.1&   &  &  &  \vspace{4pt}\\
  TERZAN 7 & $ 37.4 \pm  3.4 $ & $ 17.6 \pm  2.2 $ &       14.0&  &  &  &  \vspace{4pt}\\
\textbf{TERZAN 8} & $ 16.6 \pm  2.2 $ & $ 11.8 \pm  1.8 $ & \textbf{3.4}&  &  &  &  \vspace{4pt}\\
    \hline
    \end{tabular}
  
    \label{tab:bin_frac}
\end{table*}

\section{Initial binary fraction comparison}
\label{secA:fbin_comparison}

\begin{figure*}
    \centering
    \includegraphics[width=0.8\linewidth]{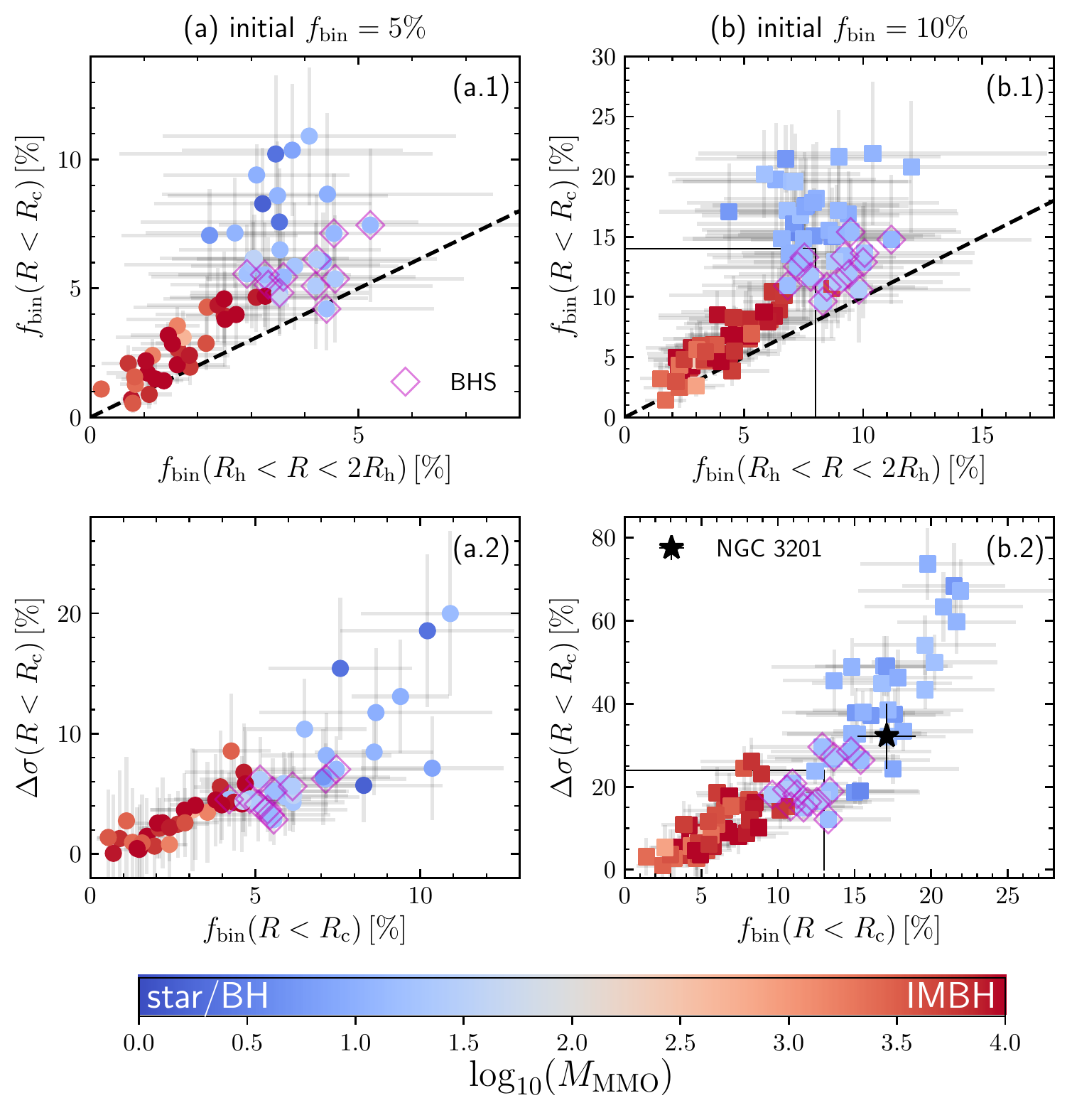}
    \caption{Comparison between a sample of simulated GCs with initial $f_{\text{bin}}=5\%$ and GCs with initial $f_{\text{bin}}=10\%$. Panels (a.1) and (b.1) show the binary fractions within the cluster core and around the half-light radius for the cases with $f_{\text{bin}}=5\%$ and $f_{\text{bin}}=10\%$, respectively (as in Figure \protect\ref{fig:imbh_bin_frac}). Panels (a.2) and (b.2) show the velocity dispersion difference for both samples (as in Figure \protect\ref{fig:imbh_bin_kin}). Both samples show similar trends, where GCs hosting a central IMBH have lower binary fractions and lower velocity dispersion differences. We can notice too that, while their behaviours are similar, the sample with initial $f_{\text{bin}}=5\%$ is scaled down to lower binary fractions. For clarity, in panels (b.1) and (b.2), the black lines delimit the region shown in panels (a.1) and (a.2). Note that panel (b.1) corresponds to Figure \protect\ref{fig:imbh_bin_frac} and panel (b.2) corresponds to the bottom panel of Figure \protect\ref{fig:imbh_bin_kin}.}
    \label{fig:init_fbin_5_10}
\end{figure*}

In addition to our sample with initial $f_{\text{bin}}=10\%$, we analyzed a sample of 65 GCs with initial $f_{\text{bin}}=5\%$. Figure \ref{fig:init_fbin_5_10} shows the binary fractions (as in Figure \ref{fig:imbh_bin_frac}) and velocity dispersion differences (as in Figure \ref{fig:imbh_bin_kin}) for both samples. We can see that the general behaviour of the parameters is similar between the two samples. However, for the sample with initial $f_{\text{bin}}=5\%$, all points are scaled-down, consistently with the initially lower binary fraction. Furthermore, as shown by the black lines in panels (b.1) and (b.2) in Figure \ref{fig:init_fbin_5_10}, all clusters with initial  $f_{\text{bin}}=5\%$ cover the same region in binary fraction space as the GCs hosting an IMBH in the sample with initial $f_{\text{bin}}=10\%$. We expect that simulations with initial $f_{\text{bin}}>10\%$ will show the same behaviour but scaled up to larger binary fractions. 

To fairly compare GCs with different initial binary fractions, finding a normalisation scheme for both binary fraction and velocity dispersion is necessary. While this escapes the aim of this work, it will be crucial for further implementing the analysis shown here.

%%%%%%%%%%%%%%%%%%%%%%%%%%%%%%%%%%%%%%%%%%%%%%%%%%

% Don't change these lines
\bsp	% typesetting comment
\label{lastpage}
\end{document}